\documentclass[modern,trackchanges]{aastex631}
\usepackage{ulem,multirow}

\begin{document}

\title{Dependence of the Radio Emission on the Eddington Ratio of
  Radio-Quiet Quasars} 
\author[0000-0003-2688-7191]{Abdulla Alhosani}
\affiliation{NYU Abu Dhabi, PO Box 129188, Abu Dhabi, UAE}
\email{ama1029@nyu.edu}

\author[0000-0003-4679-1058]{Joseph D. Gelfand}
\affiliation{NYU Abu Dhabi, PO Box 129188, Abu Dhabi, UAE}
\email{jg168@nyu.edu}

\author[0000-0002-5208-1426]{Ingyin Zaw}
\affiliation{NYU Abu Dhabi, PO Box 129188, Abu Dhabi, UAE}
\email{iz6@nyu.edu}

\author[0000-0002-1615-179X]{Ari Laor}
\affiliation{Technion Israel Institute of Technology, Department of
  Physics, Haifa 3200003, Israel} 
\email{laor@physics.technion.ac.il}

\author[0000-0002-9356-1645]{Ehud Behar}
\affiliation{Technion Israel Institute of Technology, Department of
  Physics, Haifa 3200003, Israel} 
\email{behar@physics.technion.ac.il}

\author[0000-0003-1586-3653]{Sina Chen}
\affiliation{Technion Israel Institute of Technology, Department of
  Physics, Haifa 3200003, Israel} 
\email{sina.chen@campus.technion.ac.il}

\author[0000-0002-2200-0592]{Ramon Wrzosek}
\affiliation{Rice University, Department of Physics and Astronomy,
  P.O. Box 1892, Houston, Texas 77251-1892} 
\email{ramon.wrzosek@rice.edu}

\begin{abstract}
Roughly $10\%$ of quasars are “radio-loud”, producing copious radio
emission in large jets.  The origin of the low-level radio emission
seen from the remaining $90\%$ of quasars is unclear. Observing a
sample of eight radio-quiet quasars with the Very Long Baseline Array,
we discovered that their radio properties depend strongly on their
Eddington ratio $r_{\rm Edd} \equiv L_{\rm AGN}/L_{\rm Edd}$. At lower
Eddington ratios $r_{\rm Edd} \lesssim 0.3$, the total radio emission
of the AGN predominately originates from an extremely compact region,
possibly as small as the accretion disk. At higher Eddington ratios
($r_{\rm Edd} \gtrsim 0.3$), the relative contribution of this compact
region decreases significantly, and though the total radio power
remains about the same, the emission now originates from regions
$\gtrsim100~{\rm pc}$ large. The change in the physical origin of the
radio-emitting plasma region with $r_{\rm Edd}$ is unexpected, as the
properties of radio-loud quasars show no dependence with Eddington
ratio. Our results suggest that at lower Eddington ratios the
magnetised plasma is likely confined by the accretion disk corona, and
only at higher Eddington ratios escapes to larger scales. Stellar-mass
black holes show a similar dependence of their radio properties on the
accretion rate, supporting the paradigm which unifies the accretion
onto black holes across the mass range. 
\end{abstract}


\section{Introduction}\label{sec1}
Quasars are the most persistent luminous sources in the Universe.
Powered by material accreting onto the super-massive black hole (SMBH)
residing at the center of a galaxy (e.g., \citealt{salpeter64,
  lynden-bell69}), the observed manifestations of these active
galactic nuclei (AGN) vary significantly across the electromagnetic
spectrum.  At radio wavelengths, there is a well established dichotomy
in the observed properties of quasars, with $\sim10\%$ being
``radio-loud", whose ratio of 6~cm (4.8~GHz) to 4400\AA~flux density
$>$ 10 \citep{1989AJ.....98.1195K}, with the remaining $\sim90\%$
being ``radio-quiet" (e.g., \citealt{sandage65} and references
therein). For radio-loud AGN, the radio emission is generated by a
relativistic jet that originates close to the event horizon of the
black hole (see \citealt{blandford19} for a recent review), regardless
of its appearance in other wavebands (e.g. \citealt{urry95} for
references thereafter).  However, the situation is very different for
radio-quiet AGN where, not only is there a multitude of possible
sources for their radio emission (see \citealt{panessa19} for a recent
review), the dominant mechanism can, and almost certainly does, vary
between different radio-quiet quasars (RQQs). Understanding how
$\sim90\%$ of accreting SMBHs produce their radio emission is
important for understanding the physics of accretion onto black
holes. 

Recent work suggests that the radio spectrum of RQQs depends on the
AGN's Eddington ratio $r_{\rm Edd} \equiv L_{\rm AGN}/L_{\rm Edd}$,
where $L_{\rm AGN}$ is its bolometric luminosity and $L_{\rm Edd}$ is
the ``maximum" (Eddington) luminosity of material accreting onto a
SMBH with its particular mass. A Very Large Array (VLA) study of the
4.8 and 8.5~GHz emission from 25 radio-quiet Palomar-Green (PG)
quasars found that RQQs with low Eddington ratios ($r_{\rm Edd}
\lesssim 0.3$) had flat radio spectra (spectral index $\alpha \gtrsim
-0.5$; flux density $S_\nu \propto \nu^\alpha$), while RQQs with
higher Eddington ratios ($r_{\rm Edd} \gtrsim 0.3$) had steep ($\alpha
\lesssim -0.5$) radio spectra \citep{laor19}.  Understanding the
physical implications of this correlation requires determining the
origin of the radio emission in these galaxies.   

In general, the radio emission from an AGN is primarily produced by
the accretion disk itself or an accretion-powered outflow (e.g. wind
or weak jet).  Radio emission from optically thick plasma -- possibly
the disk corona and/or base of a jet located very close (less than a
few parsecs) to the SMBH -- will result in a compact, flat spectrum
($\alpha\sim0$) radio source. Conversely, outflows are observed to
produce steep spectrum ($\alpha \lesssim -0.5$) radio emission on
scales ranging from parsecs (pc) to kilo-parsecs (kpc) in size.
Disentangling the emission from these two components requires
measuring the radio spectrum and morphology of an AGN on physical
scales far smaller than the $\sim0.1~{\rm kpc}$ spatial resolution of
the VLA radio data used by \citealt{laor19}.   

To rectify this situation, we studied the pc-scale emission of the PG
RQQs studied by \citealt{laor19} with the four highest radio spectral
indices and the four lowest radio spectral indices, listed in Table
\ref{Extended Table 1}.  In Section \ref{sec11} we describe our
reduction and analysis of 1.4 GHz and 4.8 GHz Very Long Baseline Array
(VLBA) observations of these PG RQQs, while in Section
\ref{sec:measurements} we present the properties of the nuclear radio
emission derived from these observations.  In Section
\ref{sec:survey}, we describe the dependence of their observed radio
properties with the Eddington ratio, while in Section
\ref{sec:conclusions} we discuss the implications of these results. 

\begin{table}[tb]
\caption{Properties of PG Radio-Quiet Quasars observed with the VLBA}
\begin{center}
\begin{tabular}{@{}lccccccc@{}}
\hline 
\hline
Name & $\alpha_{\rm J2000}$ & $\delta_{\rm J2000}$ & z & $r_{\rm Edd}$
& $\alpha_{4.8-8.5}^{\rm VLA}$ & $S_{4.8}^{\rm VLA}$ [mJy] & Ref.\\ 
(1) & (2) & (3) & (4) & (5) & (6) & (7) & (8) \\
\hline
PG 0050+124 & 00:53:34.940 & +12:41:36.20 & 0.060 & 1.07 & $-$1.45 &
$2.41 \pm 0.12$ &
\tablenotemark{a},\tablenotemark{b},\tablenotemark{c} \\ 
PG 0052+251 & 00:54:52.120 & +25:25:39.00 & 0.155 & +0.93 & +0.21 &
$0.68 \pm 0.04$ & \tablenotemark{a},\tablenotemark{d} \\
PG 1149-110 & 11:52:03.540 & -11:22:24.30 & 0.050 & 0.20 & +0.48 &
$2.27 \pm 0.05$ & \tablenotemark{a},\tablenotemark{d} \\
PG 1440+356 & 14:42:07.463 & +35:26:22.92 & 0.077 & 2.70 & $-$1.88 &
$1.24 \pm 0.07$ & \tablenotemark{a},\tablenotemark{b} \\ 
PG 1612+261 & 16:14:13.203 & +26:04:16.20 & 0.131 & 0.39 & $-$1.57 &
$5.58 \pm 0.08$ &
\tablenotemark{a},\tablenotemark{b},\tablenotemark{d} \\ 
PG 1613+658 & 16:13:57.179 & +65:43:09.58 & 0.139 & 0.08 & $+$1.06 &
$3.03 \pm 0.07$ & \tablenotemark{a}\\ 
PG 2130+099 & 21:32:27.813 & +10:08:19.46 & 0.062 & 0.85 & $-$1.40 &
$2.18 \pm 0.07$ &
\tablenotemark{a},\tablenotemark{b},\tablenotemark{d} \\  
PG 2304+042 & 23:07:02.912 & +04:35:57.22 & 0.042 & 0.03 & $+$0.67 &
$0.77 \pm 0.07$ & \tablenotemark{a}\\ 
\hline
\hline
\end{tabular}
\end{center}
\tablenotetext{a}{\cite{1989AJ.....98.1195K}} 
\tablenotetext{b}{\cite{1996AJ....111.1431B}} 
\tablenotetext{c}{\cite{1998MNRAS.297..366K}} 
\tablenotetext{d}{\cite{2006A&A...455..161L}}
\tablecomments{(1): Name  of galaxy in Palomar-Green Catalog
  \cite{1986ApJS...61..305G}. (2)(3): Right Ascension $\alpha_{\rm
    J2000}$ and Declination $\delta_{\rm J2000}$, respectively, of the
  PG RQQs. (4): z, red shift. Columns 2,3, and 4 were obtained from
  NED. The NASA/IPAC Extragalactic Database (NED) is operated by the
  Jet Propulsion Laboratory, California Institute of Technology, under
  contract with the National Aeronautics and Space
  Administration. (5): The Eddington ratio, as calculated by
  \cite{laor19}. (6) The spectral index between the 4.8 and 8.5 GHz
  peak flux density measured in VLA observations, as calculated by
  \cite{laor19}. (7): The average 4.8 GHz total flux density measured
  in past VLA observations of the targets. (8): References for the VLA
  4.8 GHz flux densities.} 
\label{Extended Table 1}
\end{table}

\section{VLBA Data Reduction}
\label{sec11}
In this section, we describe the reduction of the 1.4 and 4.8 GHz Very
Long Baseline Array (VLBA) observations of the eight Palomar-Green
Radio-Quiet Quasars (PG RQQs) listed in Table \ref{Extended Table
  1}. The properties of these VLBA observations are provided in Table
\ref{Extended Table 2}, which used all ten main VLBA stations,
recording the data from each using the Mark6 VLBI system
\citep{2013PASP..125..196W} with 2-bit sampling at a rate of 4 Gbps,
and then correlated at the remote Array Operation Center. At both 1.4
and 4.8 GHz, we used four 128 MHz, dual polarization, intermediate
frequency (IF) bands. For the 1.4 GHz datasets, these IFs were
centered at 1376, 1504, 1632, and 1760 MHz. For 4.8 GHz datasets, the
IFs were centered at 4612, 4740, 4868, and 4996 MHz. Each observation
was $\sim$3 hours long, and all began and ended with two-minute scans
of a strong calibrator source (``Fringe Finder"; Table \ref{Extended
  Table 2}) at each observing frequency.  Between the Fringe Finder
scans, we alternated between 30 second scans of the phase calibrator
and two-minute scans of the RQQ target at 1.4 GHz, and switched to 4.8
GHz halfway through the observation time.  As shown in Table
\ref{Extended Table 2}, for each target the phase calibrator was
located $\sim0.\!\!^{\circ}4 - 1.\!\!^{\circ}9$ away, less than the
$\sim5^\circ$ away as required for phase-referencing to work. This
observing strategy yielded $\sim1$ hour of total integration time on
the target at each frequency.

\begin{table}[tb]
\caption{Properties of VLBA Observations}
\begin{center}
\begin{tabular}{@{}lccccc@{}}
\toprule
Name & Observation & Fringe & Phase & Angular & Integration \\
 & Date & Finder & Calibrator & Separation & Time [mins] \\
(1) & (2) & (3) & (4) & (5) & (6) \\
\hline
PG 0050+124 & 2020 May 05 & 3C454.3 & J0055+1408 & $1.\!\!^{\circ}5$ &
58.7 , 58.7 \\ 
PG 0052+251 & 2020 May 16 & 3C454.3 & J0054+2550 & $0.\!\!^{\circ}4$ &
58.8 , 58.6  \\ 
PG 1149-110 & 2020 May 31 & 4C39.25 & J1153-1105 & $0.\!\!^{\circ}4$ &
58.6 , 58.6 \\ 
PG 1440+356 & 2020 Jun. 12 & 3C345 & J1438+3710 & $1.\!\!^{\circ}9$ &
58.6 , 58.6 \\ 
PG 1612+261 & 2020 May 12 & 3C345 & J1610+2414 & $2.\!\!^{\circ}0$ &
58.8 , 58.7 \\ 
PG 1613+658 & 2020 May 11 & J1642+3948 & J1623+6624 &
$1.\!\!^{\circ}2$ & 58.7 , 58.7 \\ 
PG 2130+099 & 2020 May 05 & J2005+7752 & J2130+0843 &
$1.\!\!^{\circ}5$ & 58.6 , 58.7 \\ 
PG 2304+042 & 2020 May 07 & 3C454.3 & J2300+0337 & $1.\!\!^{\circ}9$ &
58.9 , 58.7 \\ 
\botrule
\end{tabular}
\end{center}
\tablecomments{(1): PG name obtained from the Palomar-Green Catalog
  \citep{1986ApJS...61..305G}. (2): Date of observation. (3): Source
  used to find the fringes and calibrate the antennae bandpass and
  gains. (4): Source used to calibrate the measured phase and
  amplitude. Both calibrator sources were selected from the NRAO VLBA
  calibrator database. (5): Angular separation between the PG galaxy
  and phase calibrator (6): On-source integration time at 1.4 (left)
  and 4.8 GHz (right).} 
\label{Extended Table 2}
\end{table}

The data was edited and calibrated using the Astronomical Image
Processing System (AIPS; \citealt{wells85, greisen90}).  We first
corrected the recorded visibilities for ionospheric delays (VLBATECR)
and errors in the Earth Orientation Parameters (VLBAEOPS), and then
corrected for amplitude errors resulting from digital sampling
(VLBACCOR). We then used observations of the fringe finder to solve
for the antenna phase delays (VLBAMPCL) and calculate the bandpass of
each antenna (VLBABPSS).  Furthermore, we inspected the data on all
baselines (pairs of antennae), IF, and sources for abnormalities
which, once identified were removed.  At 1.4 GHz, this necessitated a
careful inspection of the visibilities for radio frequency
interference (RFI), which were removed from the data using the AIPS
task (SPFLG). The FRING task was then used to find the group delay and
phase rate which maximized the fringes observed from first, the fringe
finder, and then both the fringe finder and phase calibrator.  At 4.8
GHz, these initial solutions were improved by upon by
``self-calibrating" data obtained on the phase calibrator.  At 1.4
GHz, self calibration did not lead to a significant improvement in
image quality, due to the increased atmospheric coherence timescale at
this longer frequency, and therefore was not performed.  Upon
completion of the various calibration steps mentioned above, we used
the AIPS task CLCAL to calculate the final calibration table for the
target, which were then applied to the data using the AIPS task
SPLIT. 

\begin{figure}[tbh]
\minipage{0.02\textwidth}
    \raisebox{0in}{\footnotesize\rotatebox[origin=t]{90}{Low Eddington}}
\endminipage\hfill
\minipage{0.95\textwidth}
    \minipage{0.03\textwidth}
    \endminipage\hfill
    \minipage{0.24\textwidth}
        \centering
        {\footnotesize PG 2304+042}
    \endminipage\hfill
    \minipage{0.24\textwidth}
        \centering
        {\footnotesize PG 1613+658}
    \endminipage\hfill
    \minipage{0.24\textwidth}
        \centering
        {\footnotesize PG 1149-110}
    \endminipage\hfill
    \minipage{0.24\textwidth}
        \centering
        {\footnotesize PG 0052+251}
    \endminipage

    \minipage{0.03\textwidth}
        \raisebox{0in}{\footnotesize\rotatebox[origin=t]{90}{1.4 GHz}}
    \endminipage\hfill
    \minipage{0.24\textwidth}
        \centering
        \includegraphics[width=\linewidth]{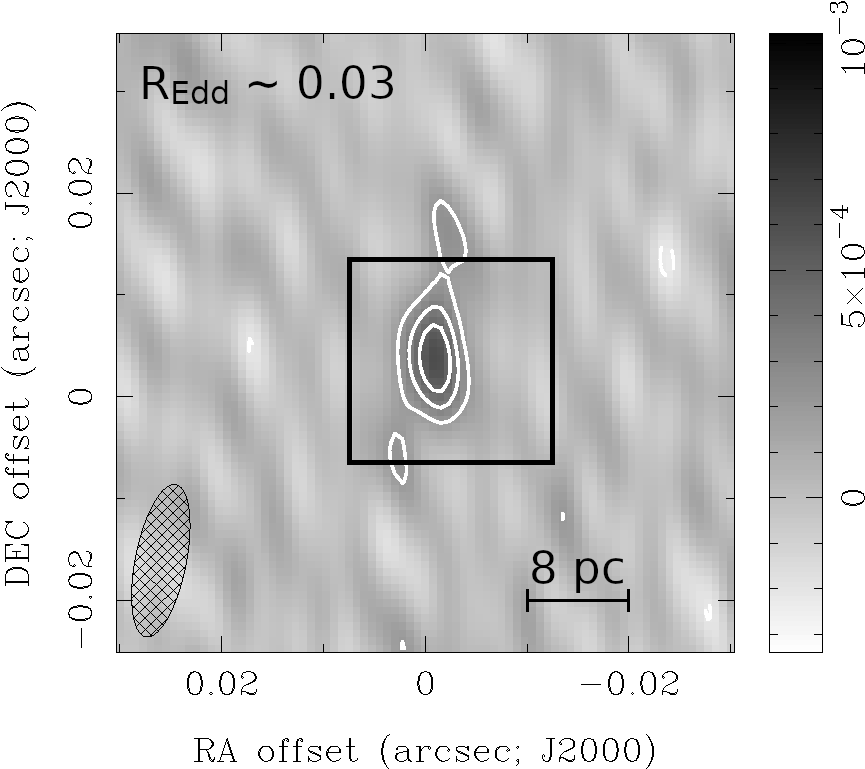}
    \endminipage\hfill
    \minipage{0.24\textwidth}
        \centering
        \includegraphics[width=\linewidth]{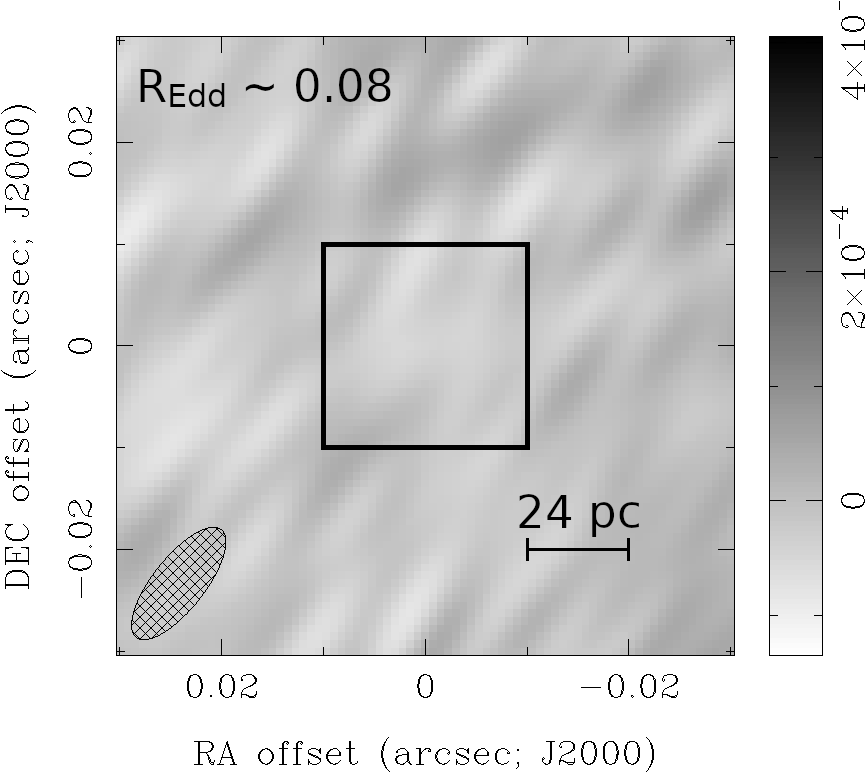}
    \endminipage\hfill
    \minipage{0.24\textwidth}
        \centering
        \includegraphics[width=\linewidth]{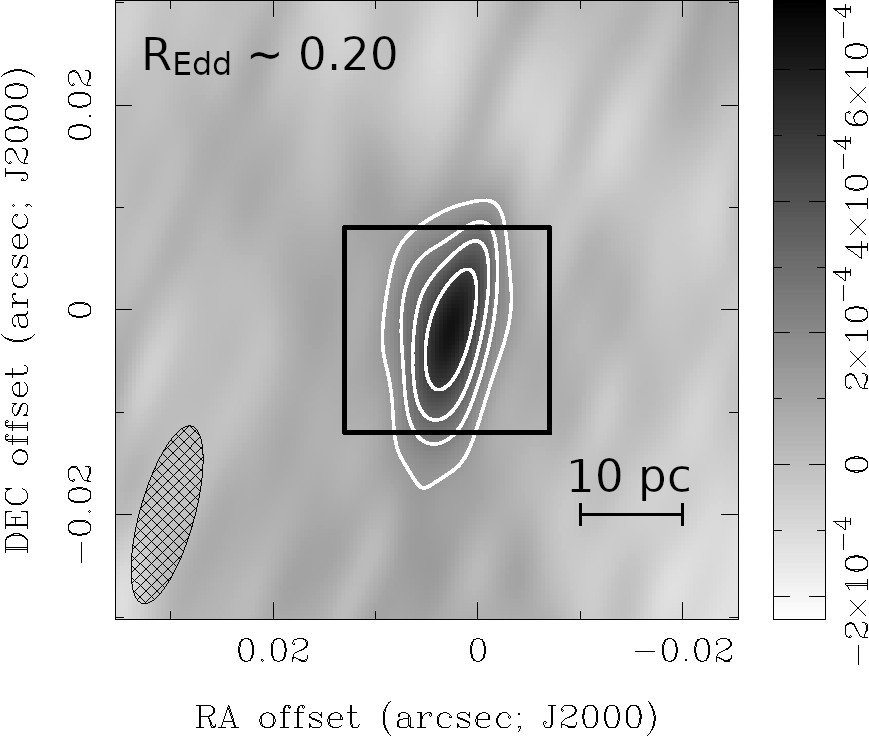}
    \endminipage\hfill
    \minipage{0.24\textwidth}
        \centering
        \includegraphics[width=\linewidth]{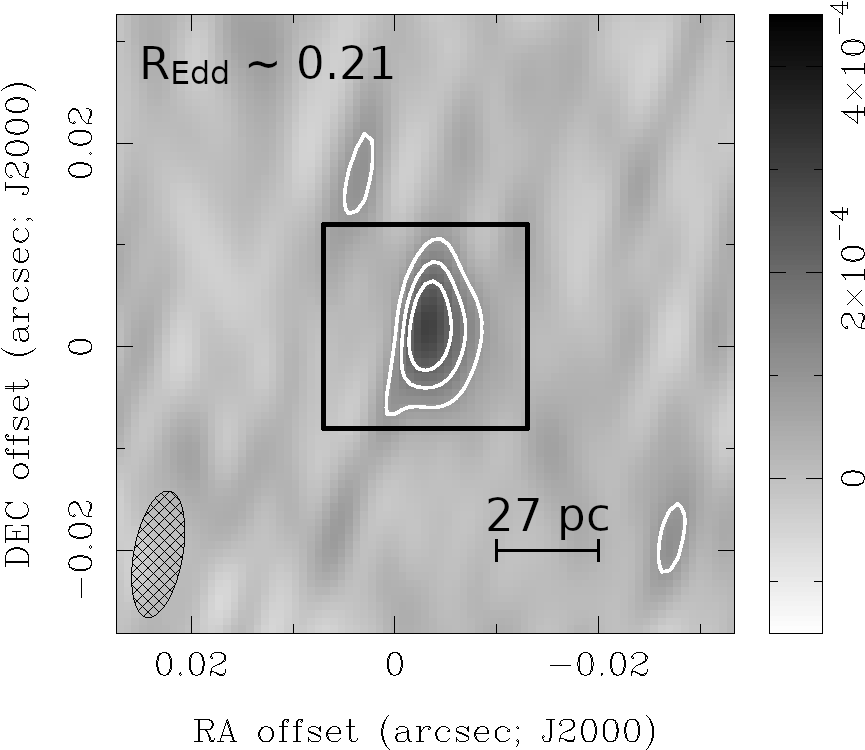}
    \endminipage

    \minipage{0.03\textwidth}
        \raisebox{0in}{\footnotesize\rotatebox[origin=t]{90}{4.8 GHz}}
    \endminipage\hfill
    \minipage{0.24\textwidth}
        \includegraphics[width=\linewidth]{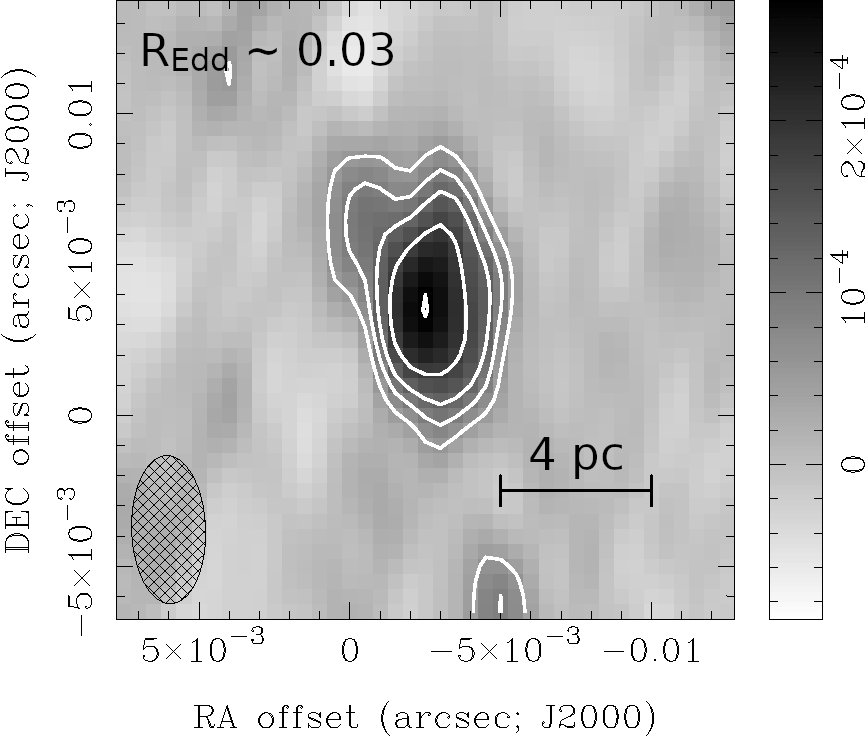}
    \endminipage\hfill
    \minipage{0.24\textwidth}
        \includegraphics[width=\linewidth]{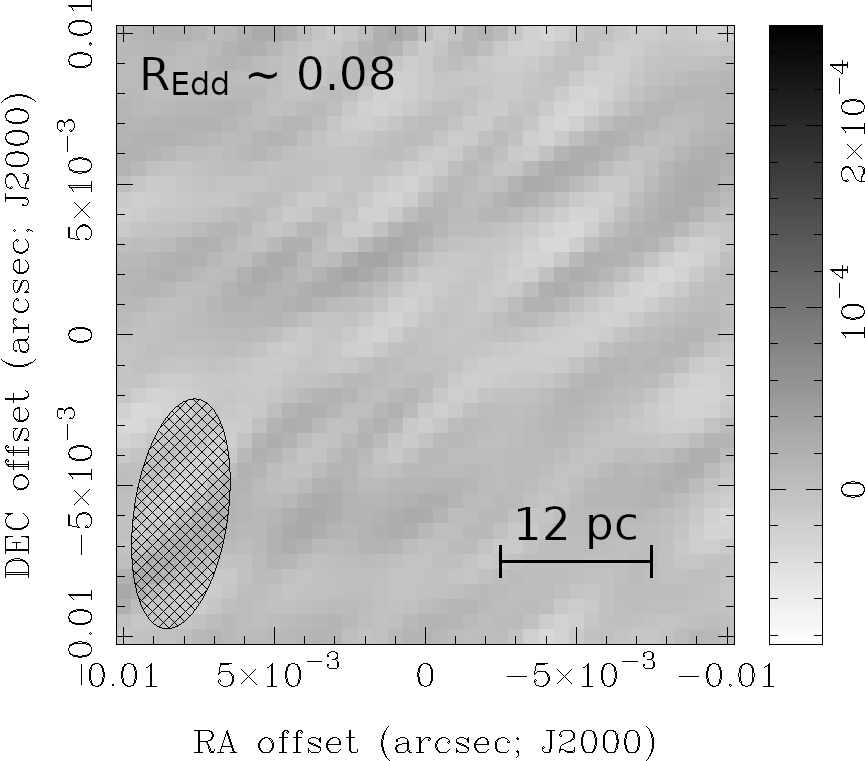}
    \endminipage\hfill
    \minipage{0.24\textwidth}
        \includegraphics[width=\linewidth]{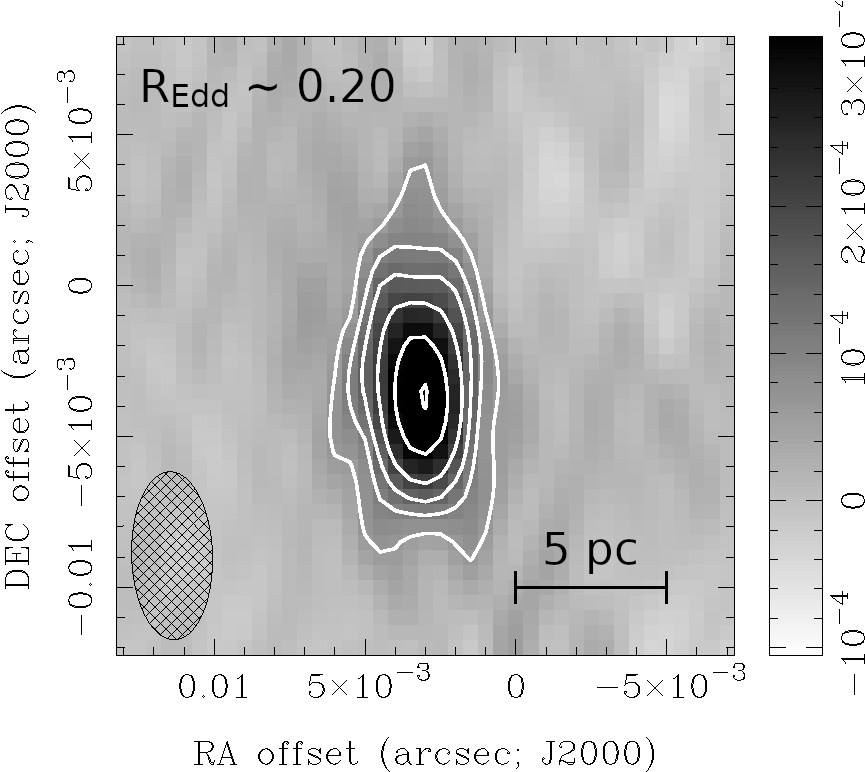}
    \endminipage\hfill
    \minipage{0.24\textwidth}
        \includegraphics[width=\linewidth]{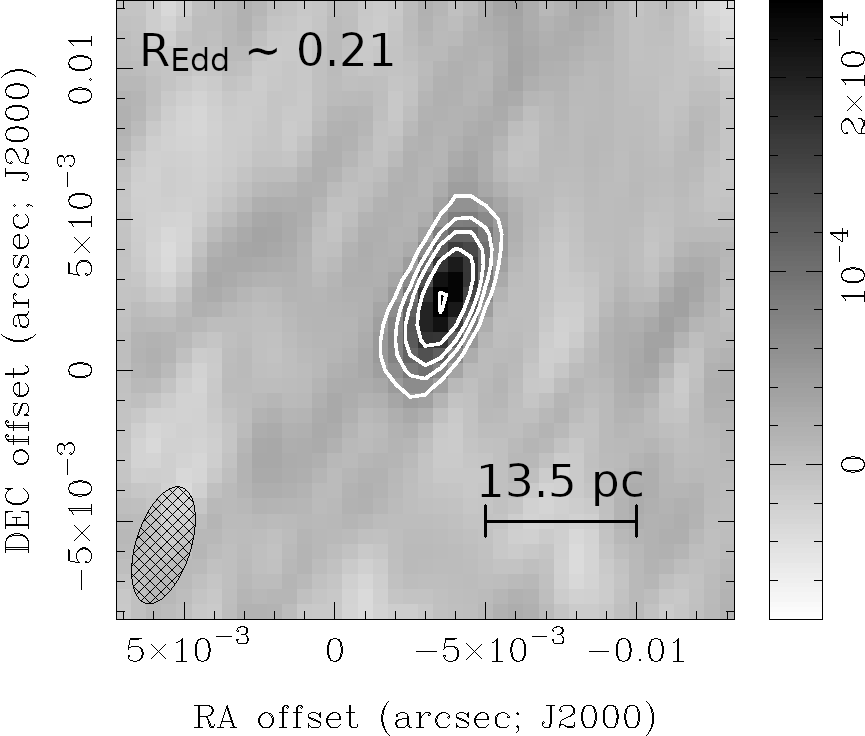}
    \endminipage
\endminipage\hfill

\vspace{0.4cm}

\minipage{0.02\textwidth}
    \raisebox{0in}{\footnotesize\rotatebox[origin=t]{90}{High Eddington}}
\endminipage\hfill
\minipage{0.95\textwidth}
    \minipage{0.03\textwidth}
    \endminipage\hfill
    \minipage{0.24\textwidth}
    \centering
    {\footnotesize PG 1612+261}
    \endminipage
    \minipage{0.24\textwidth}
    \centering
    {\footnotesize PG 2130+099}
    \endminipage
    \minipage{0.24\textwidth}
    \centering
    {\footnotesize PG 0050+124}
    \endminipage
    \minipage{0.24\textwidth}
    \centering
    {\footnotesize PG 1440+356}
    \endminipage
    
    \minipage{0.03\textwidth}
        \raisebox{0in}{\footnotesize\rotatebox[origin=t]{90}{1.4 GHz}}
    \endminipage\hfill
    \minipage{0.24\textwidth}
    \centering
      \includegraphics[width=\linewidth]{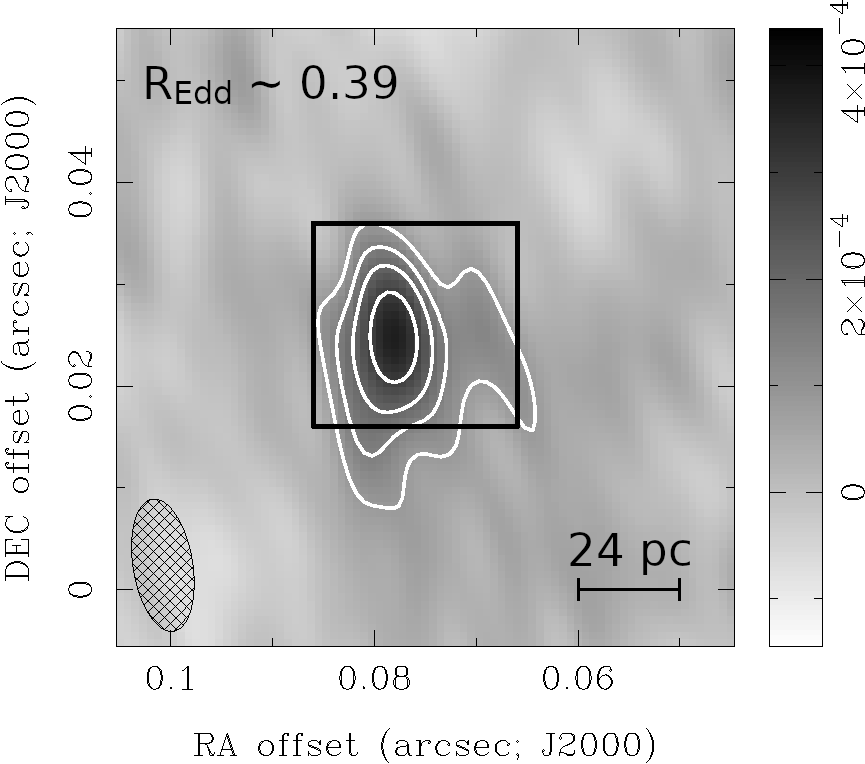}
    \endminipage
    \minipage{0.24\textwidth}
    \centering
      \includegraphics[width=\linewidth]{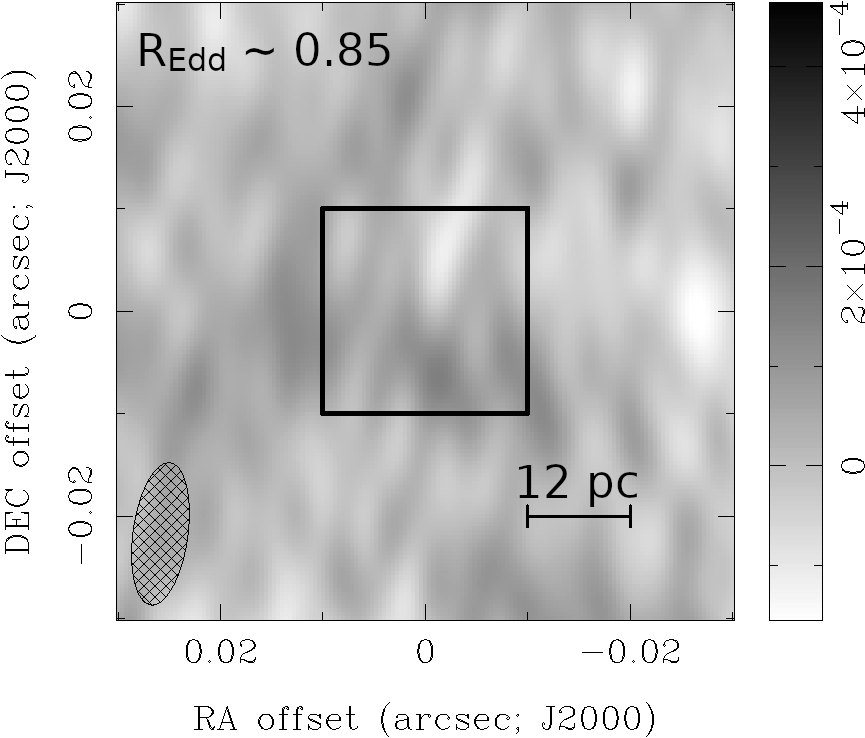}
    \endminipage
    \minipage{0.24\textwidth}
    \centering
      \includegraphics[width=\linewidth]{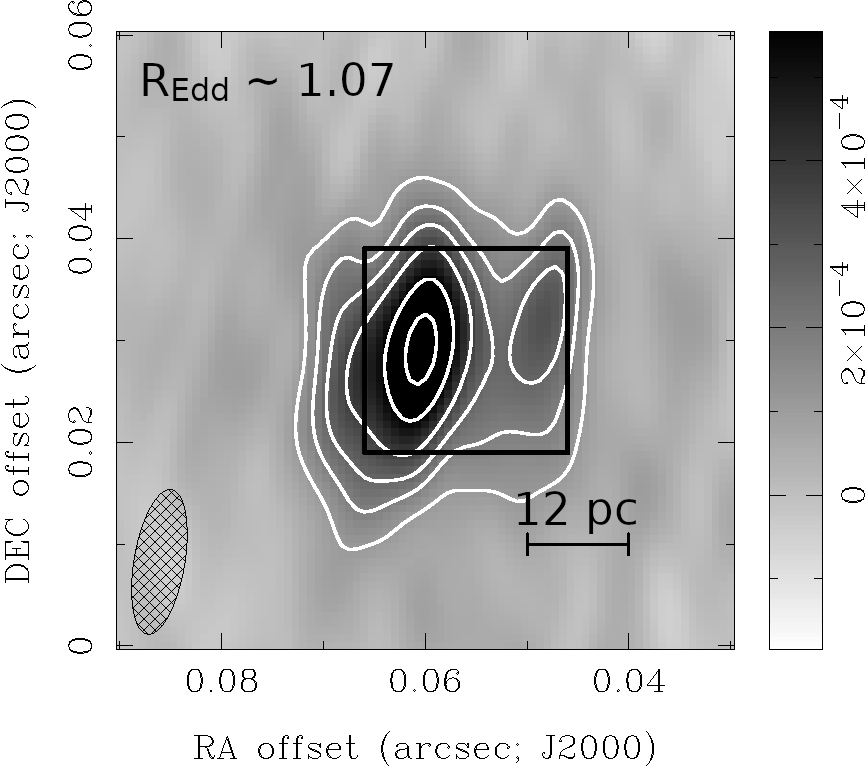}
    \endminipage
    \minipage{0.24\textwidth}
    \centering
      \includegraphics[width=\linewidth]{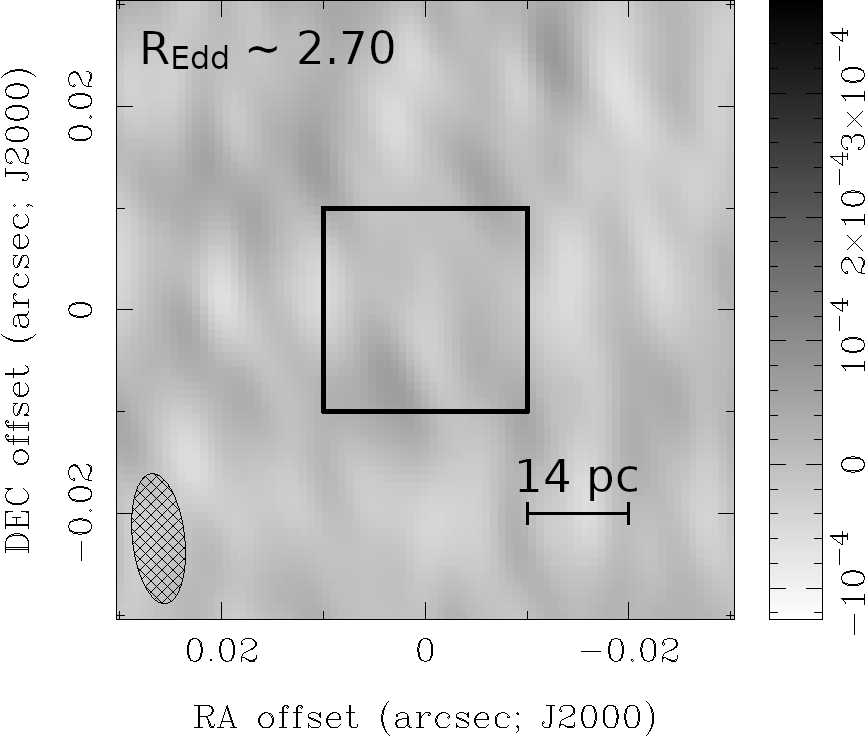}
    \endminipage
    
    \minipage{0.03\textwidth}
        \raisebox{0in}{\footnotesize\rotatebox[origin=t]{90}{4.8 GHz}}
    \endminipage\hfill
    \minipage{0.24\textwidth}
      \includegraphics[width=\linewidth]{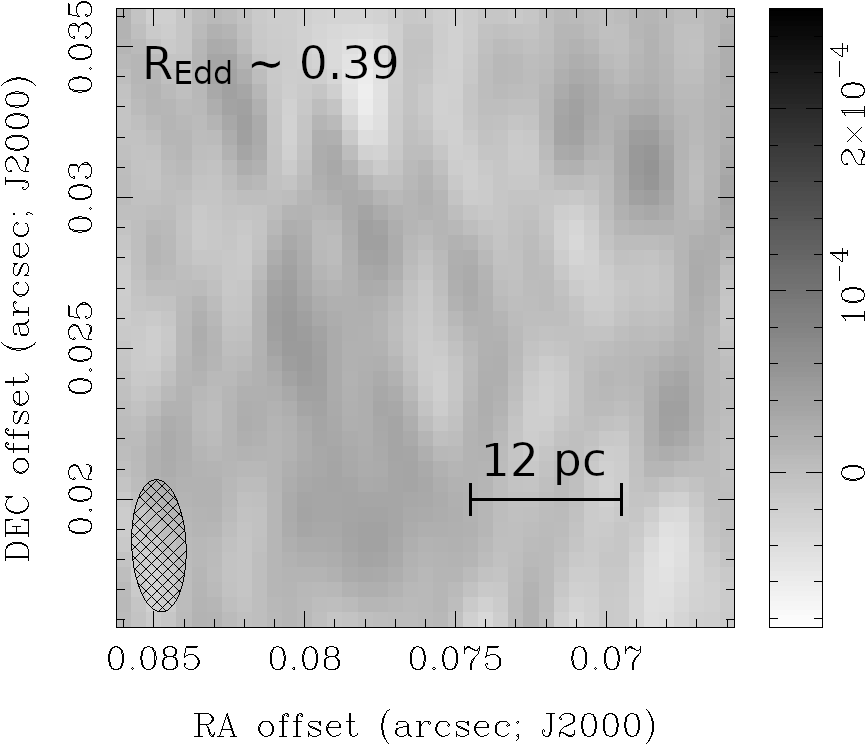}
    \endminipage
    \minipage{0.24\textwidth}
      \includegraphics[width=\linewidth]{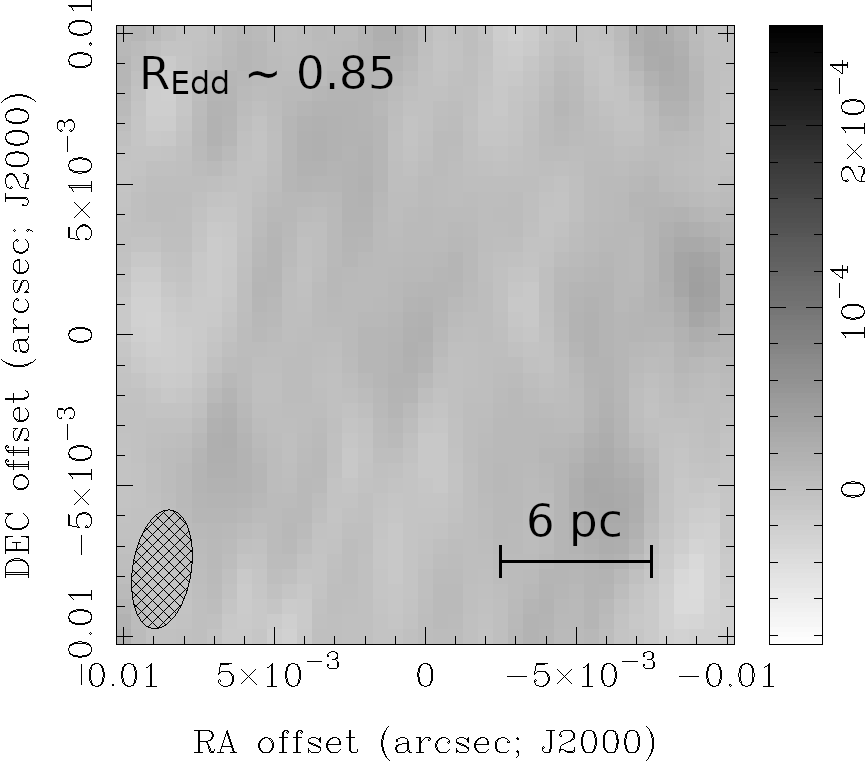}
    \endminipage
    \minipage{0.24\textwidth}
      \includegraphics[width=\linewidth]{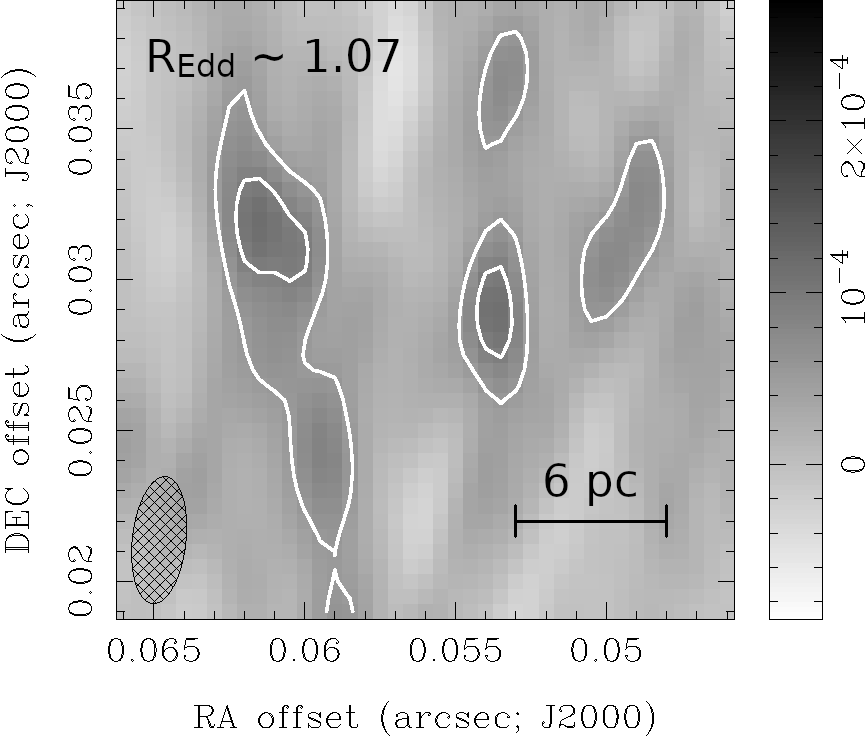}
    \endminipage
    \minipage{0.24\textwidth}
      \includegraphics[width=\linewidth]{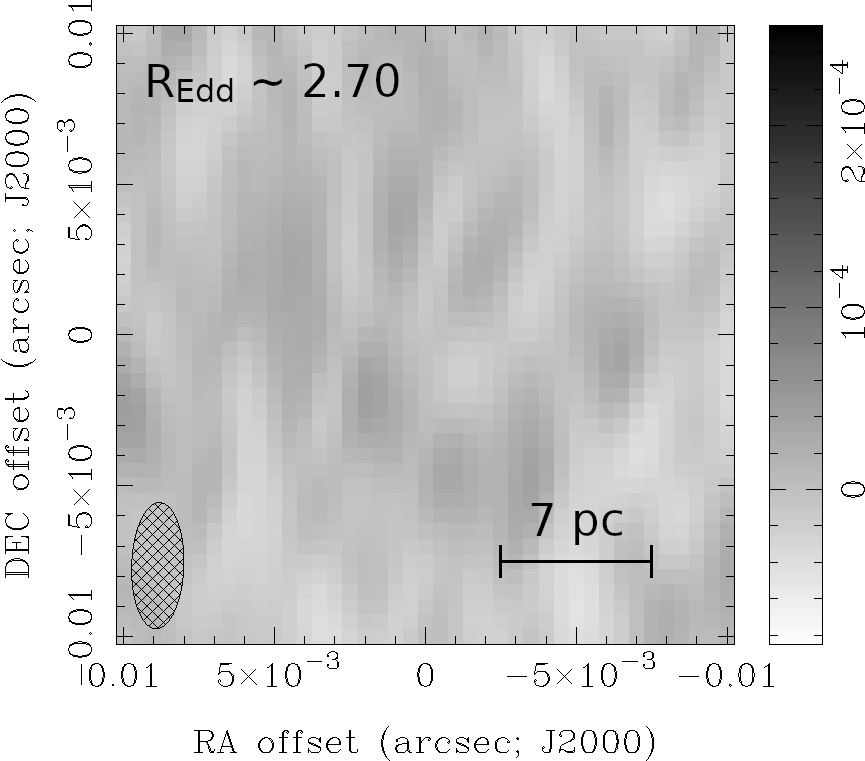}
    \endminipage
\endminipage\hfill
\caption{VLBA images of the observed PG RQQs (Table \ref{Extended
    Table 1}). For all targets, the first row is the observed VLBA
  emission at 1.4 GHz, the second row is the observed VLBA emission at
  4.8 GHz.  In each image, the origin $(0,0)$ of the coordinate system
  is the optical center of the galaxy as given in the NASA/IPAC
  Extragalactic Database (NED), the size and orientation of the
  synthesized beam is shown in the lower-left corner, the Eddington
  ratio $r_{\rm Edd}$ is indicated in the upper-left corner of the
  image, a physical size scale is indicated in the lower-right corner,
  the grey-scale indicates the intensity (in Jy/beam) ranging from
  $-5\sigma - 15\sigma$ (where the image $\sigma$ is given in Table
  \ref{Extended Table 2}), and the white contours indicate $3, 5, 7,
  10, 15, 20\sigma$ intensity emission for detections.  The smaller
  region of the 4.8 GHz image is indicated by the black square in the
  1.4 GHz image.  The low Eddington RQQs are detected at a higher rate
  and with higher significance than the high Eddington RQQs,
  especially at 4.8 GHz.}
\label{figure:1}
\end{figure}

Once calibrated, we imaged the recorded visibility data at each
frequency  using the AIPS task IMAGR.  This program uses the CLEAN
algorithm \citep{hogbom74, clark80} to deconvolve the ``dirty image"
(produced from taking the 2D Fourier Transform of the visibility) data
with the point spread function (PSF; ``dirty beam").  This requires
choosing the relative weight of data collected from different
baselines, and we chose natural weighting (i.e. ``robust"=5;
\citealt{briggs95}) which maximizes sensitivity at the expense of
angular resolution. The resultant images, which have a spatial
resolution of $\sim5-20~{\rm pc}$, are shown in Figure \ref{figure:1}
while their properties are given in Table \ref{Extended Table 3}. 

\section{Properties of Nuclear Radio Emission}
\label{sec:measurements}
As described in \S\ref{sec1}, distinguishing between radio emission
produced by an accretion disk and an accretion-powered outflow
requires measuring the spectrum and morphology of the pc-scale
emission probed by the VLBA observations discussed in \S\ref{sec11}.
Below, we describe the criteria needed for determining if a PG RQQ was
detected in these images (\S\ref{sec:detection}) and, if so, measuring
the flux density and angular distribution of this emission
(\S\ref{sec:modeling}). 

\begin{table}[tb]
\caption{Properties of the VLBA Images of the Observed PG RQQs}
\begin{center}
\begin{tabular}{@{}lccccccc@{}}
\toprule
\multirow{3}{*}{Name} & \multirow{3}{*}{$r_{\rm Edd}$}& & 1.4 GHz
$\left[\frac{\rm \mu Jy}{\rm beam} \right]$ & & & 4.8 GHz
$\left[\frac{\rm \mu Jy}{\rm beam} \right]$ & \\ 
 & & $\sigma$ & $I_{\rm max}$ & $I_{\rm min}$ & $\sigma$ & $I_{\rm
  max}$ & $I_{\rm min}$ \\ 
(1) & (2) & (3) & (4) & (5) & (6) & (7) & (8)\\
\hline
\textbf{PG 2304+042} & \textbf{0.03} & \textbf{68} & \textbf{613} &
\textbf{-260} & \textbf{18} & \textbf{273} & \textbf{-91} \\ 
PG 1613+658 & 0.08 & 27 & 133 & -134 & 17 & 74  & -73 \\
\textbf{PG 1149-110} & \textbf{0.20} & \textbf{47} & \textbf{629} &
\textbf{-122} & \textbf{21} & \textbf{425} & \textbf{-83} \\ 
\textbf{PG 0052+251} & \textbf{0.21} & \textbf{30} & \textbf{297} &
\textbf{-102} & \textbf{16} & \textbf{244} & \textbf{-70} \\ 
\hline
\hline
\textbf{PG 1612+261} & \textbf{0.39} & \textbf{29} & \textbf{366} &
\textbf{-139} & 17 & 75 & -78 \\ 
PG 2130+099 & 0.85 & 31 & 151 & -135 & 15 & 80 & -70 \\
\textbf{PG 0050+124} & \textbf{1.07} & \textbf{37} & \textbf{813} &
\textbf{-118} & 17 & 91 & -75 \\ 
PG 1440+356 & 2.70 & 25 & 108 & -111 & 17 & 72 & -70 \\
\botrule
\end{tabular}
\end{center}
\tablecomments{(1): PG name of target galaxy
  \citep{1986ApJS...61..305G}, listed in order of increasing Eddington
  ratio $r_{\rm Edd}$ (2). (3)(6): The root-mean-squared ($\sigma$) of
  pixel values in the 1.4 and 4.8 GHz images, respectively. (4)(7):
  The maximum pixel value $I_{\rm max}$ for the 1.4 and 4.8 GHz
  images. (5)(8):The minimum pixel value $I_{\rm min}$ for the 1.4 and
  4.8 GHz images.  The rows in bold indicate a source was detected in
  the image using the criteria defined in \S\ref{sec:detection}.} 
\label{Extended Table 3}
\end{table}

\subsection{VLBA Detections}
\label{sec:detection}
If an image contains no emission from an astronomical source, the
distribution of its pixel intensities should be well described by a
Gaussian with an average much smaller than the root-mean-squared
($\sigma$) intensity.  The symmetric nature of this distribution
results in comparable absolute values for the maximum $I_{\rm max}$
and minimum $I_{\rm min}$ pixel intensities ("brightness"), though
these values will likely be significantly larger than the noise level
due to the large number of pixels\footnote{The 1.4 GHz images have
  $2048\times2048$ pixels, while the 4.8 GHz images have
  $1024\times1024$ pixels.} in each image ($I_{\rm max} \approx
|I_{\rm min}| \gg \sigma$).  However, an image where emission from a
source is detected will have $I_{\rm max} > |I_{\rm min}| > \sigma$,
and therefore we require a detection to have a peak intensity $I_{\rm
  max}\geq5\sigma$ and $I_{\rm max} \geq 2|I_{\rm min}|$. As shown in
Table \ref{Extended Table 3}, these criteria indicate that emission is
detected in the 1.4 GHz VLBA images of five RQQs, three of which are
also detected in their 4.8 GHz VLBA images. 

\begin{figure}
\minipage{0.5\textwidth}
\centering
1.4 GHz
   \includegraphics[width=\linewidth]{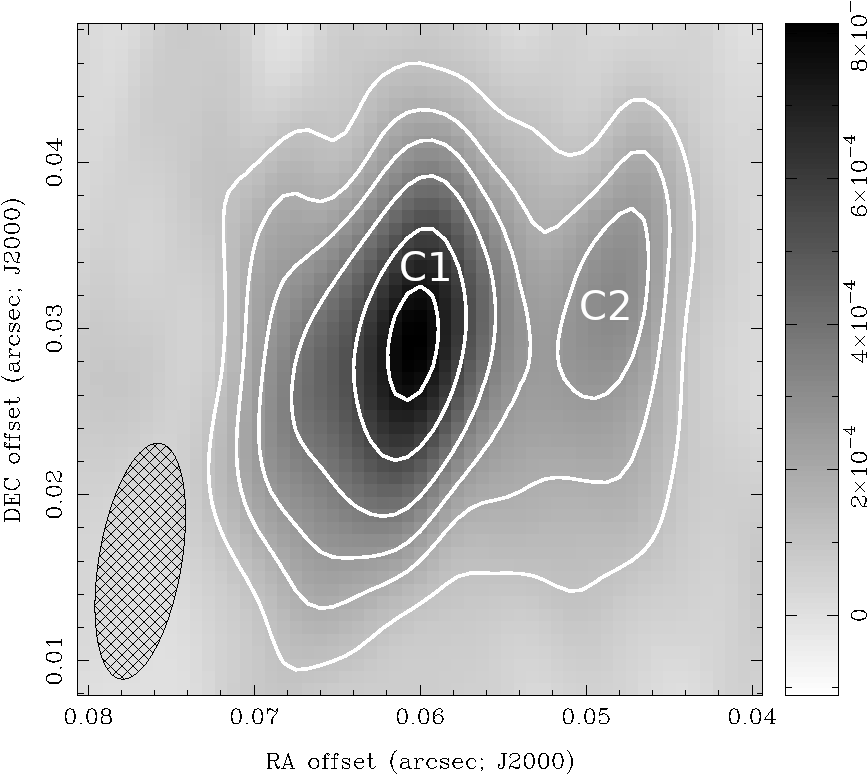}
\endminipage\hfill
\minipage{0.5\textwidth}
\centering
4.8 GHz
    \includegraphics[width=\linewidth]{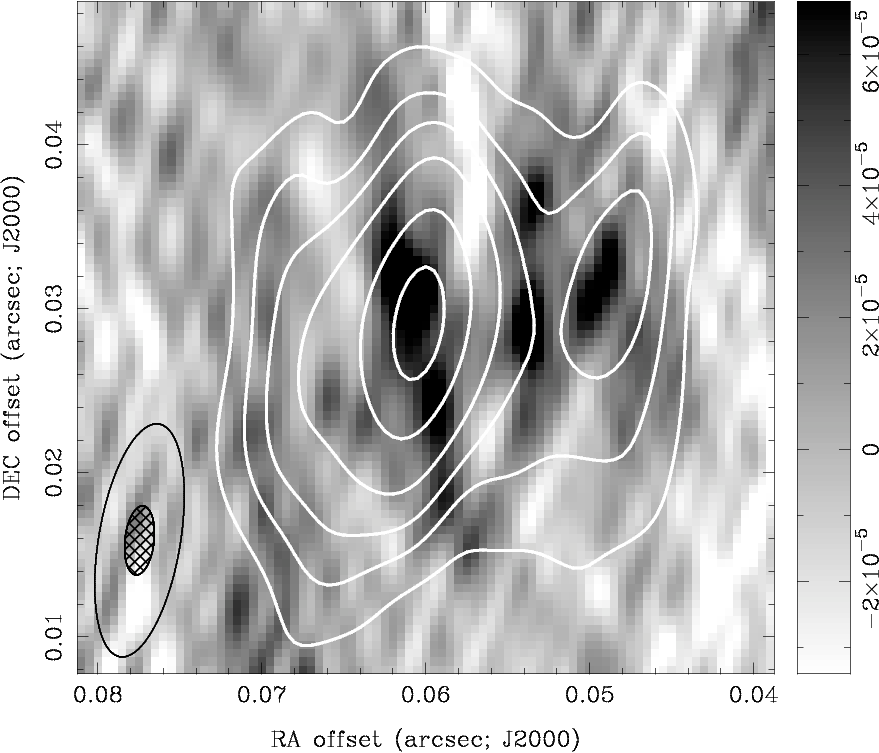}
\endminipage\hfill
\caption{40 mas $\times$ 40 mas VLBA images of the 1.4 GHz ({\it
    left}) and 4.8 GHz ({\it right}) emission of PG 0050+124. The
  white contours in both images indicate 1.4 GHz flux density
  3,5,7,10,15, and 20$\sigma$ of the 1.4 GHz image.  In the left
  image, the cross-hatched ellipse represents the beam, while in the
  right image the larger ellipse is the size of the 1.4 GHz beam while
  cross-hatched ellipse is the size of the 4.8 GHz beam.} 
\label{multi_comp}
\end{figure}

Since PG 1612$+$261 and PG 0050$+$124 were only detected at 1.4 GHz
according these criteria, we examined their 4.8 GHz images for
emission spatially coincident with that observed at the lower
frequency.  No such emission was detected in the 4.8 GHz image of PG
1612+261 (Fig.\ \ref{figure:1}), but a $\sim4-5\sigma$ excess was
detected in the 4.8 GHz image of PG 0050+124 (Fig.\ \ref{multi_comp}).
While this does not the satisfy the detection criteria described
above, the statistical significance of this spatial coincidence
suggests it is 4.8 GHz emission from this RQQ, and is treated as such
below.  

\subsection{Properties of Detected Nuclear Radio Emission}
\label{sec:modeling}
For the RQQs detected in our VLBA images, we used the AIPS task JMFIT
to fit the pixel intensities in the region around the image peak
intensity the image with a 2D Gaussian model, whose free parameters
are the centroid location, peak intensity, major and minor axis, and
orientation (position angle).  JMFIT then calculates the integrated
flux density using these values and the size of the synthesized beam
as determined by the task IMAGR.  A single 2D Gaussian was sufficient
to model the observed emission for all RQQ detections but PG 0050+124,
which required two separate Gaussian components at both 1.4 and 4.8
GHz (Fig.\ \ref{multi_comp}).

\subsubsection{Position of Nuclear Radio Emission}
\label{sec:position}
\begin{table}[tb]
    \caption{Positional Offsets and Uncertainties of Nuclear Radio Emission}
    \begin{center}
    \begin{tabular}{cccccc}
    \hline
    \hline
        \multirow{3}{*}{Name} & \multirow{3}{*}{$r_{\rm Edd}$} &
        \multicolumn{2}{c}{JMFIT Offsets} & \multicolumn{2}{c}{Optical
          position} \\ 
         &  & $\Delta_{\alpha \cos \delta}$ & $\Delta_\delta$ &
        $\sigma_{\rm pos}$ & \multirow{2}{*}{Citation} \\ 
         & & [mas] & [mas] & [mas] & \\
        (1) & (2) & (3) & (4) & (5) & (6) \\
    \hline
         PG 2304+042 & 0.03 & -2.6 & +3.7 & $\sim500$ &
         \citet{klemola87} \\ 
         PG 1149$-$110 & 0.20 & +3.2 & -3.4 & $\sim500$ & B. Skiff
         (private comm. to NED) \\ 
         PG 0052+251 & 0.21 & -3.6 & +2.4 & $\sim500$ & B. Skiff
         (private comm. to NED) \\ 
         \hline
         \hline
         PG 1612+261 & 0.39 & +83 & +25 & $\sim500$ & \citet{sdssdr6}\\
         PG 0050+124-C1 & \multirow{2}{*}{1.07} & +61 & +31 &
         \multirow{2}{*}{$\sim250$} &
         \multirow{2}{*}{\citet{clements81}} \\ 
         PG 0050+124-C2 &  & +54 & +29 \\
    \hline
    \hline
    \end{tabular}
    \end{center}
    \tablecomments{(1): PG name of target galaxy
      \cite{1986ApJS...61..305G} (2): Eddington ratio of target galaxy
      \cite{laor19}. (3)(4): Offset between the centroid of VLBA
      emission, as determined using JMFIT, from the optical center of
      the PG galaxy as provided in NED (Table \ref{Extended Table
        1}). For all but PG 1612+261, this was done using their 4.8
      GHz images. The statistical errors in this fits are all
      $\sim0.1~{\rm mas}.$ (5) $1\sigma$ uncertainty in the optical
      position of the galaxy. (6) Reference for uncertainty in optical
      position.} 
    \label{tab:vlba_pos}
\end{table}

As listed in Table \ref{tab:vlba_pos}, the radio emission detected in
our VLBA observations appears displaced from the optical center of
these galaxies, with the positions derived from image fitting
described above suggesting offsets ranging from $\lesssim5~{\rm mas}$
to $\sim70-100~{\rm mas}$ (Table \ref{tab:vlba_pos}).  Such an offset
could result from errors in the absolute astrometry of our VLBA
images.  However, using the results of a study conducted by
\cite{pradel06} to determine the positional uncertainty arising from
the ``phase referencing'' technique described in \S\ref{sec11}, we
obtain errors $\lesssim0.25~{\rm mas}$ -- significantly smaller than
what is observed.  However, this observed offset could also result
from the uncertainty in the optical position used for the phase center
of our VLBA observations, which were taken from the NASA/IPAC
Extragalactic Database (NED)\footnote{The NASA/IPAC Extragalactic
  Database (NED) is funded by the National Aeronautics and Space
  Administration and operated by the California Institute of
  Technology.}.  As listed in Table \ref{tab:vlba_pos}, this
uncertainty is considerably larger than what is observed offsets and
therefore we can not conclude if the nuclear radio emission is
displaced from the center of these galaxies as observed in other
quasars with more precise optical positions (e.g., \citealt{kovalev17,
  yao21}).  However, we note that the observed offsets
($\sim60-90$~{\rm mas}) for the higher Eddington ratio $r_{\rm Edd}
\gtrsim 0.3$ quasars are significantly larger than the offsets
($\lesssim5~{\rm mas}$) for the low Eddington ratio $r_{\rm Edd}
\lesssim 0.3$ quasars.

\subsubsection{Integrated Flux Density and Spectrum of Nuclear Radio Emission}
\label{sec:spectrum}
The resultant integrated flux densities of these components are listed
Table \ref{Extended Table 5}.  However, the measured flux densities
only include emission from the angular scales probed by the
individual pairs of antennae, or baselines, in this array.  As
described below, this has considerable implications when calculating
the spectral index $\alpha$ of a source, which requires measuring this
flux density at two frequencies. 

\begin{table}[tb]
\caption{Flux density and Spectral Index of the Detected VLBA Radio Emission}
\begin{center}
\begin{tabular}{cccccc}
\toprule
Name & $\frac{b_{\rm min}}{1000 \lambda}$ & $\theta_{\rm las} [{\rm
    mas}]$ & $S_{\rm 1.4}$ [${\rm mJy}$] & $S_{\rm 4.8}$ [${\rm mJy}$]
& $\alpha_{\rm compact}$\\ 
(1) & (2) & (3) & (4) & (5) & (6) \\
\hline
\multirow{2}{*}{PG 2304+042} & 700 & $\sim 300$ & $0.53 \pm 0.08$ &
$\cdots$ & $\cdots$ \\ 
 & 2500 & $\sim 85$ & $0.55 \pm 0.07$ & $0.49 \pm 0.04$ & $-0.09 \pm 0.12$ \\
\hline
\multirow{2}{*}{PG 1149-110} & 800 & $\sim 260$ & $1.06 \pm 0.11$ &
$\cdots$ & $\cdots$ \\ 
 & 2800 & $\sim 75$ & $0.86 \pm 0.11$ & $0.59 \pm 0.04$ & $-0.30 \pm 0.11$ \\
\hline
\multirow{2}{*}{PG 0052+251} & 860 & $\sim 240$ & $0.42 \pm 0.07$ &
$\cdots$ & $\cdots$ \\ 
 & 3000 & $\sim 70$ & $0.35 \pm 0.06$ & $0.30 \pm 0.03$ & $-0.13 \pm 0.16$ \\
\hline
\hline
\multirow{2}{*}{PG 1612+261} & 1000 & $\sim 205$ & $0.95 \pm 0.11$ &
$\cdots$ & $\cdots$ \\ 
 & 3500 & $\sim 60$ &  $0.58 \pm 0.08$ & $<0.08$ & $<-1.53$ \\
\hline
\multirow{2}{*}{PG 0050+124-C1} & 860 & $\sim 240$ & $2.66 \pm 0.15$ &
$\cdots$ & $\cdots$ \\ 
 & 3000 & $\sim 70$ & $1.42 \pm 0.10$ & $0.31 \pm 0.07$  & $-1.21 \pm 0.19$ \\
\hline
\multirow{2}{*}{PG 0050+124-C2} & 860 & $\sim 240$ & $0.72 \pm 0.12$ &
$\cdots$ & $\cdots$ \\ 
 & 3000 & $\sim 70$ & $0.24 \pm 0.07$ & $0.24 \pm 0.05$ & $0 \pm 0.29$ \\
\botrule
\end{tabular}
\end{center}
\tablecomments{(1): PG name of target, listed in order of increasing
  Eddington ratio $r_{\rm Edd}$ (2): Minimum baseline length $b_{\rm
    min}$ of the data included in the image in units of
  $1000\times$the observing wavelength $\lambda$ (equivalent to the
  ``uvrang'' parameter in IMAGR used to make these images).
  Differences between targets reflect their different positions on the
  sky during their observations. (3): The largest angular scale
  $\theta_{\rm las}$ of the resultant image, as calculated using
  Equation \ref{eqn:theta_las}. (4)(5): Integrated 1.4 and 4.8 GHz
  flux densities of the source in the corresponding image.  For
  non-detections, a $5\sigma$ upper limit is provided. (6): Spectral
  index, $\alpha_{\rm compact}$ calculated using the integrated 1.4
  and 4.8 GHz flux densities measured in images with the same largest
  angular scale.} 
\label{Extended Table 5}
\end{table}

For a baseline of (projected) length\footnote{The projected length is
  defined to be the distance between two antenna as measured from the
  viewpoint of the source.} $b$, the measured visibility is the
intensity emitted on an angular scale $\theta \sim \frac{\lambda}{b} =
\frac{c}{b\nu}$, where $\lambda$ is observed wavelength, $\nu$ is the
observed frequency, and $c$ is the speed of light.  Therefore, an
observation on the same baseline $b$ at two different frequencies
$\nu$ measures the intensity not only emitted at different frequencies
but also from different angular scales -- with observations at 4.8 GHz
($\lambda \approx 6~{\rm cm}$) measuring the intensity on scales
$\sim3.5\times$ smaller (i.e., more compact) than that measured at 1.4
GHz ($\lambda \approx 20~{\rm cm}$).  Since our 1.4 and 4.8 GHz VLBA
observations (Table \ref{Extended Table 2}) were performed using the
same array, and therefore the same distribution of baselines, the
emission detected at each frequency originates from different, but
overlapping, ranges of angular scales.  As a result, differences in
the measured 1.4 and 4.8 GHz flux density of a source not only
reflects its intrinsic spectrum (i.e., intensity as a function of
frequency $\nu$) but its morphology (i.e., intensity as a function of
angular scale $\theta$).  

Measuring the intrinsic spectral index $\alpha$ of these sources
requires measuring their flux density $S_\nu$ at different frequencies
$\nu$ but on the same range of angular scales $\theta$.  This requires
producing 1.4 and 4.8 GHz images sensitive to the same range of
angular scales.  The flux density measured from an image produced at
frequency $\nu$ using data from baselines longer than $b>b_{\rm
  min,\nu}$ only includes emission originating from angular scales
$\theta \lesssim \theta_{\rm las}$, where the largest angular scale
is: 
\begin{eqnarray}
\label{eqn:theta_las}
\theta_{\rm las} & \sim & \frac{\lambda}{b_{\rm min,\nu}} =
\frac{c}{b_{\rm min,\nu} \nu}. 
\end{eqnarray}  
Therefore, producing 1.4 and 4.8 GHz images with the same $\theta_{\rm
  las}$ requires the length of the shortest baseline used at 1.4 GHz
$(b_{\rm min,1.4})$ is $\frac{4.8}{1.4}\approx3.5\times$ longer than
the shortest baseline used at 4.8~GHz $(b_{\rm min,4.8})$. 

To maximize the data used in this analysis, we set $b_{\rm min,4.8} =
b_{\rm min}$, the projected distance of the shortest baseline during a
particular VLBA observation.  We then made a 1.4 GHz image only using
data from baselines with length $b \gtrsim 3.5b_{\rm min}$ by setting
the ``uvrange" parameter in AIPS task IMAGR (\S \ref{sec11}) to the
appropriate value (given in Table \ref{Extended Table 5}).  The flux
density of the source in this 1.4 GHz image was measured using this
same image fitting routine described above, with the resultant values
listed in Table \ref{Extended Table 5}.  We then use this 1.4 GHz flux
density to calculate the spectral index $\alpha_{\rm compact}$ (flux
density $S_\nu \propto \nu^\alpha$) of the 4.8 GHz emission detected
in an image produced using all baselines, also listed in Table
\ref{Extended Table 5}. 

\begin{table}[tb]
\caption{Deconvolved Image Sizes}
\begin{center}
\begin{tabular}{cccccccc}
\toprule
\multirow{2}{*}{Name} & \multirow{2}{*}{$r_{\rm Edd}$} & $\nu$ &
$\theta_{\rm M} \times \theta_{\rm m}$ & \multirow{2}{*}{$\frac{\rm
    pc}{\rm mas}$} & $d_{\rm M} \times d_{\rm m}$ & $A_{\rm proj}$ &
1.4 GHz \\ 
& & [GHz] & ${\rm mas}\times{\rm mas}$ & & ${\rm pc}\times{\rm pc}$ &
pc$^{2}$ & Compactness \\ 
(1) & (2) & (3) & (4) & (5) & (6) & (7) & (8) \\
\hline
PG2304+042 & 0.03 & 4.8 & $4.0\times2.3$ & 0.83 & $3.4\times1.9$ & 5 &
$1.0\pm0.2$\\ 
\hline
\multirow{2}{*}{PG1149-110} & \multirow{2}{*}{0.20} & 1.4 &
$13.8\times6.7$ & \multirow{2}{*}{0.98} & $13.6\times6.5$ & 70 &
\multirow{2}{*}{$0.8\pm0.1$}\\ 
& & 4.8 & $4.3\times1.9$ & & $4.3\times1.9$ & 6 & \\
\hline
\multirow{2}{*}{PG0052+251} & \multirow{2}{*}{0.21} & 1.4 &
$8.5\times5.3$ & \multirow{2}{*}{2.71} & $23\times14$ & 250 &
\multirow{2}{*}{$0.8\pm0.2$}\\ 
 & & 4.8 & $2.4\times0.3$ & & $6.5\times0.9$ & 4.5 & \\
\hline
\hline
PG1612+261 & 0.39 & 1.4 & 11.2$\times$7.3 & 2.35 & 26.2$\times$17.2 &
355 & $0.6\pm0.1$ \\ 
\hline
\multirow{2}{*}{PG0050+124-C1} & \multirow{4}{*}{1.07} & 1.4 &
$16.2\times11.7$ & \multirow{4}{*}{1.17} & $18.9\times13.6$ & 200 &
\multirow{2}{*}{$0.50\pm0.05$}\\ 
& & 4.8 & $10.1\times2.7$ & & $11.8\times3.2$ & 30 & \\
\multirow{2}{*}{PG0050+124-C2} &  & 1.4 & $24.9\times5.6$ &  &
$29.0\times6.5$ & 150 & \multirow{2}{*}{$0.3\pm0.1$} \\ 
 & & 4.8 & $5.7\times2.8$ & & $6.6\times3.3$ & 20 & \\
\botrule
\end{tabular}
\end{center}
\tablecomments{(1): PG name of target quasar. (2): Eddington
  ratio. (3): Frequency of image. (4): Deconvolved major $\theta_{\rm
    M}$ and minor $\theta_{\rm m}$ axis calculated by JMFIT for a
  particular image where possible. PG 2304+042 is unresolved at 1.4
  GHz, and PG 1612+261 was undetected at 4.8 GHz. (5): The number of
  pc corresponding to angular size of 1~mas and the angular distance
  to the galaxy, as calculated using \citet{wright06} for $H_0 =
  69.6~{\rm km~s^{-1}~Mpc^{-1}}$, $\Omega_m=0.286$,
  $\Omega_\Lambda=0.714$ \citep{bennett14}. (6): Major $d_{\rm M}$ and
  minor $d_{\rm m}$ axes in parsec. (7): Projected physical area
  $A_{\rm proj}$ of the source assuming it is an ellipse. (8): 1.4 GHz
  Compactness, calculated using Equation \ref{eqn:compactness} using
  the quantities given in Table \ref{Extended Table 5}. } 
\label{tab:area}
\end{table}

\subsubsection{Extent of Nuclear Radio Emission}
\label{sec:extent}
Furthermore, if the observed emission originates from a region
considerably larger than the synthesized beam, then JMFIT (\S
\ref{sec11}) returns its deconvolved angular size.  Since this
quantity accounts for the size of the synthesized beam in the image,
it should be the same at 1.4 and 4.8 GHz if the same physical region
is responsible for the emission observed at both frequencies.
However, as listed in Table \ref{tab:area}, this is not the case, with
the deconvolved size at 1.4~GHz $\sim5-50\times$ larger than at 4.8
GHz for sources resolved at both frequencies.  This discrepancy
results from the different angular scales probed by VLBA observations
at different frequencies, as discussed in \S\ref{sec:spectrum}, and
the larger size observed at 1.4 GHz indicates emission on scales
greater than detectable at 4.8 GHz.

Therefore, the deconvolved size is not the physical extent of the
emission region, and we require a different way of estimating this
quantity.  The dependence between measured flux density and baseline
length at a particular frequency allows us to quantitatively measure
the distribution of emission on different angular scales, i.e., its
morphology.  We do so by comparing the 1.4 GHz integrated flux density
measured in an image made using all VLBA baselines $S_{1.4}(>b_{\rm
  min})$ to that measured in an image produced only using the
baselines probing angular scales also measured at 4.8~GHz,
$S_{1.4}(>3.5b_{\rm min})$.  If the emitting region is infinitesimally
small, its intensity will be the same on all baselines, and therefore
$S_{1.4}(>3.5b_{\rm min}) \approx S_{1.4}(>b_{\rm min})$.  However, a
large emission region will produce a higher intensity on shorter
baselines (i.e., larger angular scales) than longer baselines (i.e.,
smaller angular scales), and therefore $S_{1.4}(>3.5b_{\rm min}) <
S_{1.4}(>b_{\rm min})$.  As a result, the ratio of these two flux
densities is inversely correlated to the extent of the emitting
region.  We therefore define the ``1.4 GHz Compactness" of a source to
be:
\begin{eqnarray}
\label{eqn:compactness}
{\rm 1.4~GHz~Compactness} & \equiv & \frac{S_{1.4}(>3.5b_{\rm
    min})}{S_{1.4}(>b_{\rm min})}, 
\end{eqnarray}
such that $0\lesssim {\rm 1.4~GHz~Compactness} \lesssim 1$, and a more
compact (higher fraction of the total emission originating from a
smaller volume) source will have a larger 1.4~GHz Compactness than a
more diffuse source.  We calculated this quantity using the flux
densities given in Table \ref{Extended Table 5}, with the results
listed in Table \ref{tab:area}.

\section{Eddington Ratio Dependence of PG RQQ Radio Properties}
\label{sec:survey}
Using the morphological and spectral properties of the radio emission
of these RQQs described in \S\ref{sec:measurements}, we can now
disentangle emission from the innermost accretion disk (compact, flat
$\alpha \sim 0$ spectrum) and an accretion-powered outflow (diffuse,
steep $\alpha \lesssim -0.5$ spectrum). In this Section, we discuss
the different origins of the radio emission observed from these eight
PG RQQs, and how their properties change with Eddington ratio $r_{\rm
  Edd}$ of these AGN. 

\begin{figure}[tb]
\begin{center}
  \includegraphics[width=0.75\linewidth]{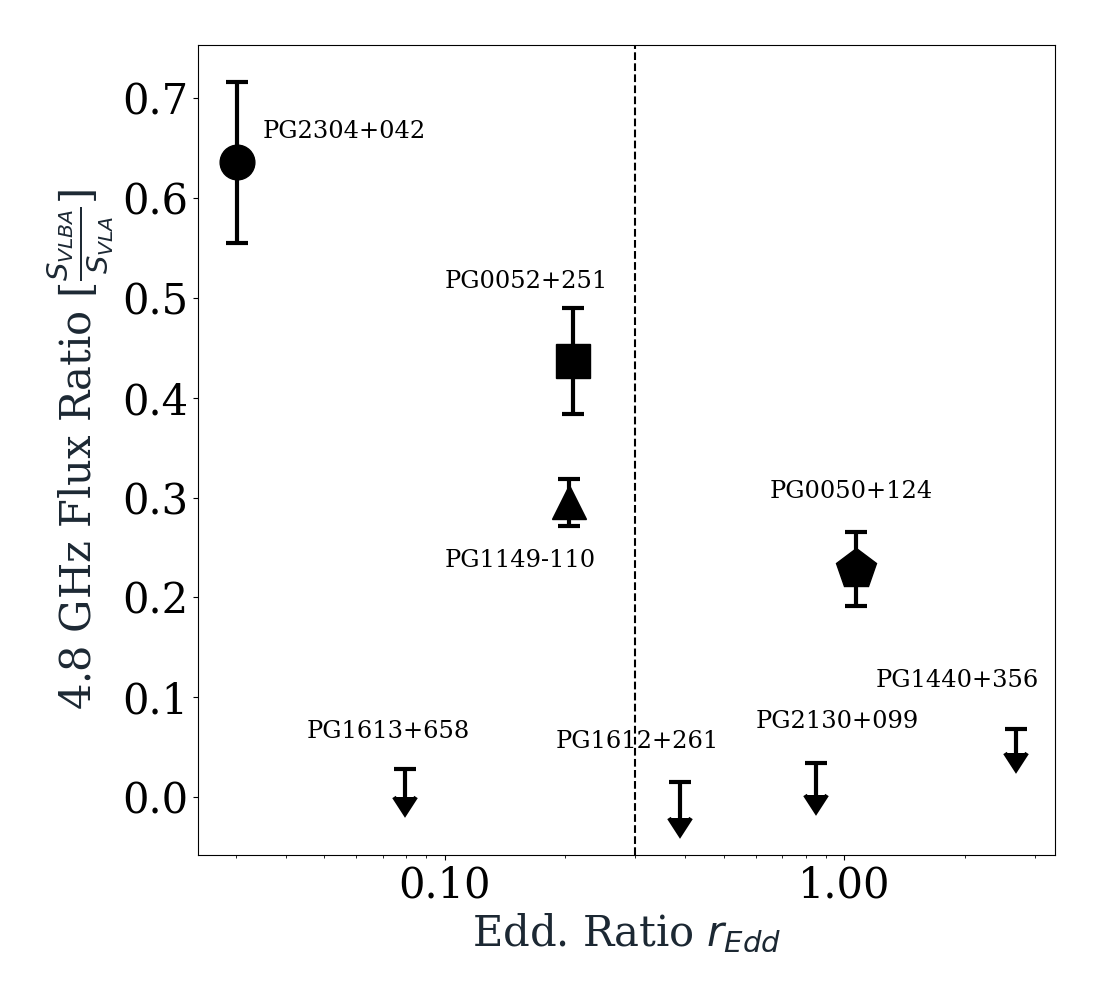}
\end{center}
\caption{Ratio of the VLBA and VLA 4.8 GHz integrated flux densities
  $[\frac{S_{VLBA}}{S_{VLA}}]$ of the surveyed PG RQQs as a function
  of their Eddington ratio $r_{\rm Edd}$. The VLA 4.8 GHz flux density
  came from archival work, listed in Table \ref{Extended Table 1}, the
  upper-limits of the VLBA 4.8 GHz flux density of the non-detections
  is $5\times$ the image $\sigma$ given in Table \ref{Extended Table
    2}, and the VLBA 4.8 GHz flux density of detected sources is given
  in Table \ref{Extended Table 5}. The vertical dashed line indicates
  an Eddington ratio $r_{\rm Edd}=0.3$ -- AGN with lower values are
  considered to have low Eddington ratios. A non-detection by the VLBA
  suggests that $\lesssim 10\%$ of the AGN's total radio emission
  comes from regions $\lesssim100~{\rm pc}$.  Furthermore, for VLBA
  detections, the fraction of the total 4.8 GHz flux density coming
  from such small regions decreases as $r_{\rm Edd}$ increases.} 
\label{fig:nuc_ratio}
\end{figure}

While our 4.8 GHz VLBA observation of these PG RQQs can only detect
emission originating in regions $\lesssim0.1~{\rm kpc}$ in size,
previous VLA observations measured the total emission from regions
$\lesssim5~{\rm kpc}$ large.  Since  these past 4.8 GHz VLA
observations measured flux densities (VLA $S_{4.8}$ Table
\ref{Extended Table 1}) greatly exceeding the noise level
$(>30\sigma)$ of our 4.8 GHz VLBA images (Table \ref{Extended Table
  2}), a VLBA non-detection of a particular RQQ indicates its 4.8 GHz
radio emission originates from a region $\gtrsim0.1~{\rm kpc}$ in
size, and/or has significantly decreased during the $\sim20-30$ years
since the VLA observation, with the short variability timescale
requiring a compact emission region.  We are unable to distinguish
between these two possibilities for the only low Eddington PG RQQ not
detected in our VLBA observations (PG 1613+658) since there exists
only one previous 4.8 GHz VLA observation of this source.  However,
all of the high Eddington ratio PQ RQQs undetected in our 4.8 GHz VLBA
observations (PG 1440+356, PG 1612+261, and PG 2130+099) were detected
with a nearly constant flux densities in multiple 4.8 GHz VLA
observations (Table \ref{Extended Table 1}), suggesting their radio
emission primarily originates from a large region and therefore
produced by an outflow.   

Similarly, the detection of 4.8 GHz VLBA emission from a PG RQQ
suggests a significant contribution from compact ($\lesssim0.1~{\rm
  kpc}$) regions.  The higher detection rate of low Eddingtion RQQs in
our 4.8 GHz VLBA observations suggests a higher fraction of their
radio emission originates from smaller regions than their high
Eddington counterparts.  To further investigate this dependence, we
calculate the ratio of a RQQ's 4.8 GHz VLBA to VLA flux densities,
which constrains the fraction of the AGN's total radio emission
originating from its nucleus (defined as the central $\lesssim0.1~{\rm
  kpc}$ of the galaxy in this work).  As shown in Figure
\ref{fig:nuc_ratio}, $\lesssim20\%$ of the total 4.8 GHz radio
emission from high Eddington ratio RQQs originates from inside their
nucleus -- indicating their radio emission is predominately produced
by a larger scale outflow. However, the nucleus of low Eddington ratio
RQQs can contribute as much as $\sim70\%$ of their total 4.8 GHz radio
emission. The increased 4.8 GHz VLBA detection rate and VLBA-to-VLA
flux ratio of low Eddington ratio RQQs suggests that nuclear radio
emission is both more common and more prominent in such AGN. 

\begin{figure}[tb]
\begin{center}
    \includegraphics[width=0.31\linewidth]{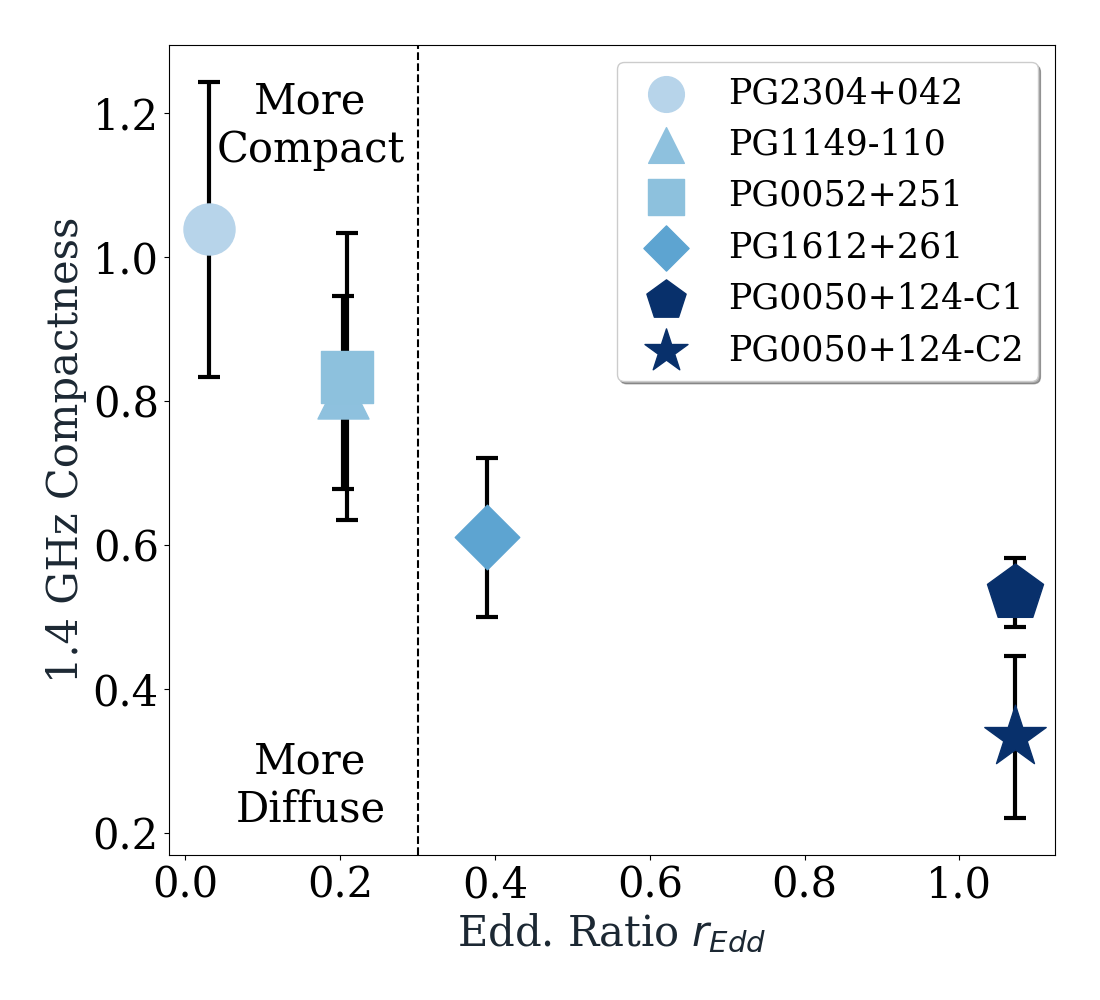}
    \includegraphics[width=0.31\linewidth]{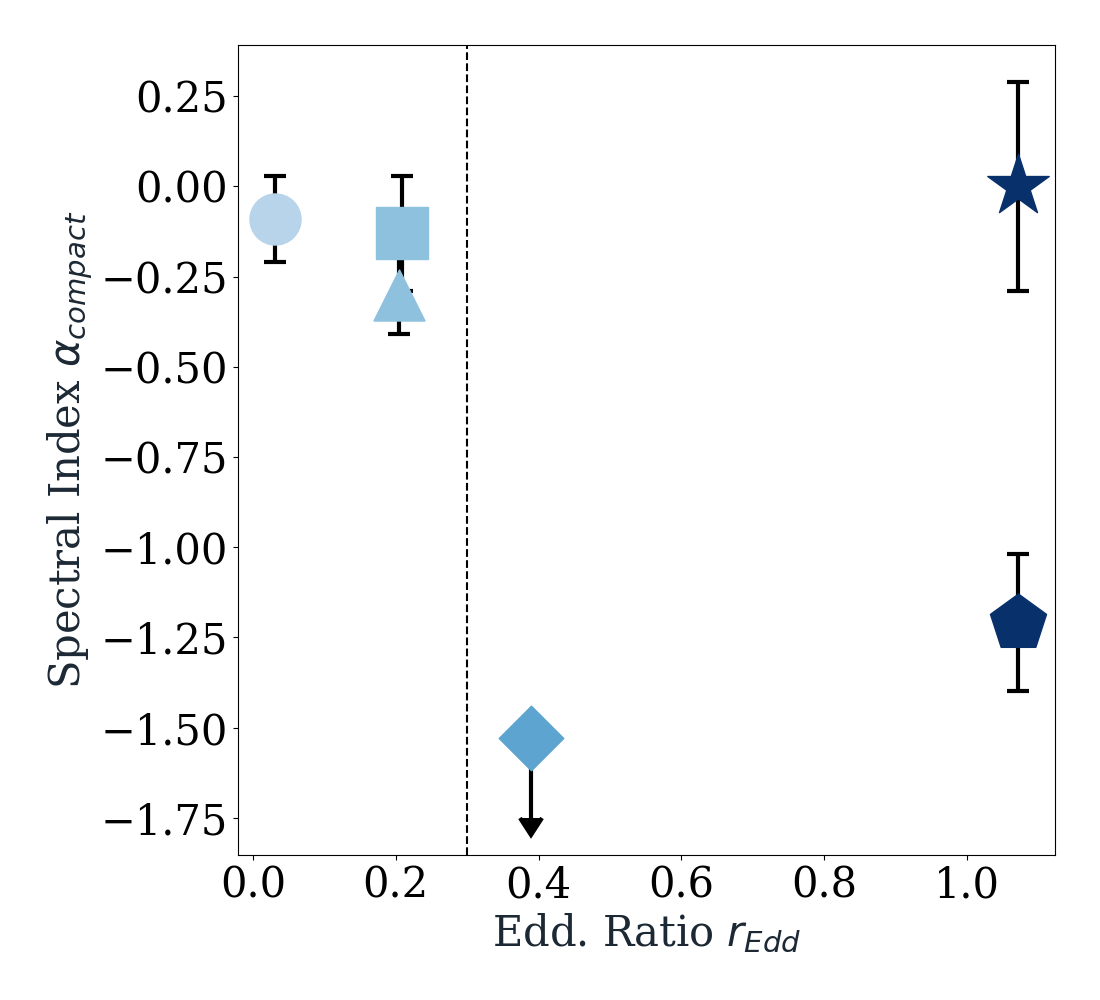}
    \includegraphics[width=0.365\linewidth]{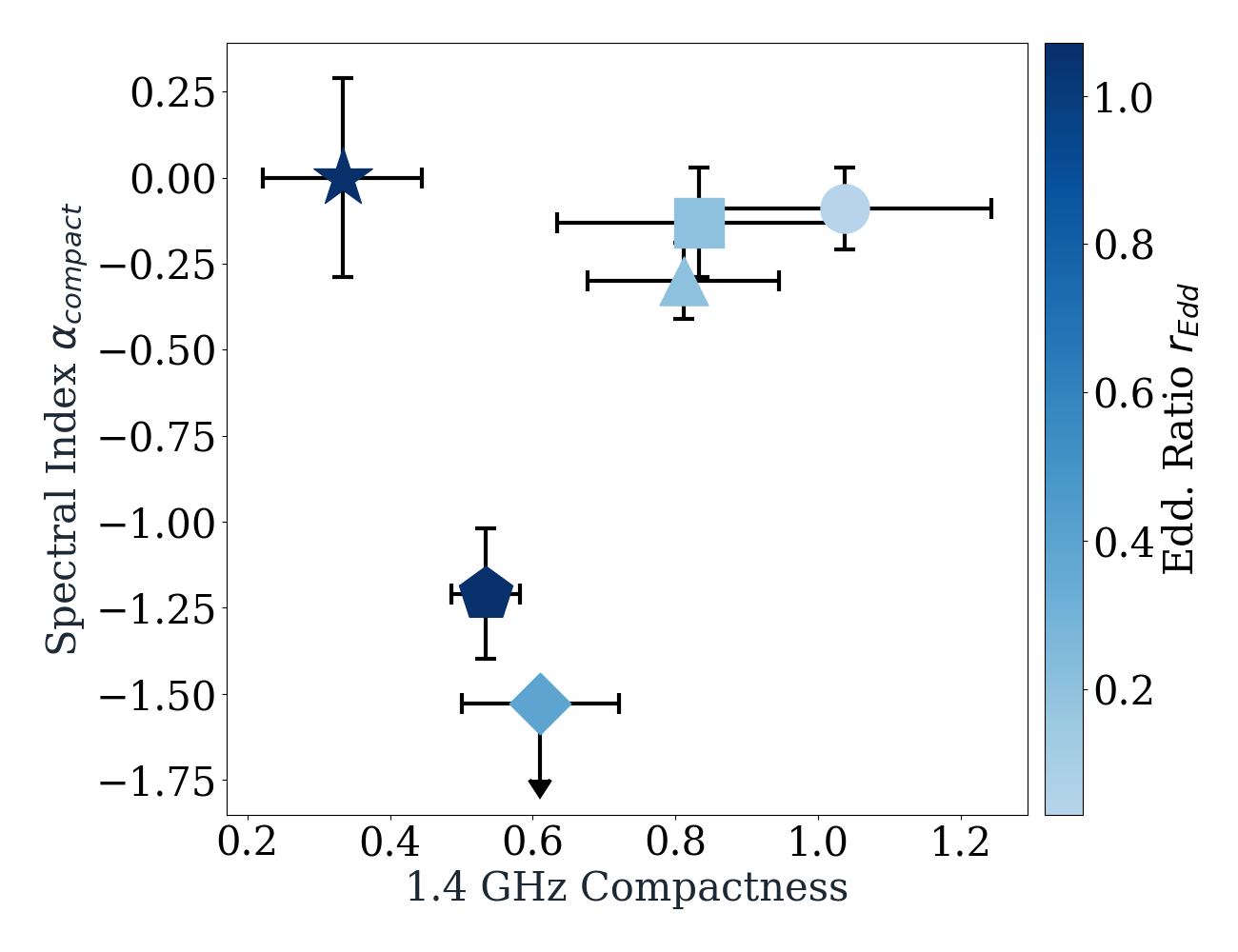}
\end{center}
\caption{{\it Left}: Compactness of the detected 1.4 GHz nuclear
  emission of these RQQs. This is calculated using the 1.4 GHz flux
  densities provided in Table \ref{Extended Table 5}, and defined in
  Equation \ref{eqn:compactness}.  As $r_{\rm Edd}$ increases, we find
  that the nuclear emission becomes less compact, i.e. more diffuse.  
{\it Middle}: Spectral index of the ``compact" nuclear emission
$\alpha_{\rm compact}$, defined to be emission from regions small
enough to be detected in our 4.8 GHz VLBA observation as described in
\S\ref{sec:modeling}, whose values are provided in Table \ref{Extended
  Table 5}. The low Eddington RQQs have flat spectra, while both high
Eddington RQQs have steep spectrum components.  In both the {\it Left}
and {\it Middle} panels, the vertical line set at $r_{\rm Edd} = 0.3$
separates low and high Eddington RQQs.  
{\it Right}: Spectral index of the compact nuclear emission
$\alpha_{\rm compact}$ vs. the 1.4 GHz compactness for the sources
detected in our VLBA images. The color bar indicates the Eddington
ratio $r_{\rm Edd}$ of the PQ RQQs, with the same color scale used for
all three plots.  For PG 0050+124, we indicate the properties of the
individual components, C1 and C2, observed in this RQQ
(Fig. \ref{multi_comp}).  We find that the radio properties of the
observed RQQs are strongly dependent with their Eddington ratio: as
the Eddington ratio of a RQQ increases, its radio nuclear emission
becomes increasingly diffuse and has a steeper radio spectrum
(Fig. \ref{VS. EDD RATIO}).  These results suggest that radio emission
from the accretion disk (compact and flat spectrum) is responsible for
a higher fraction of the nuclear radio emission in low Eddington ratio
than high Eddington ratio RQQs.} 
\label{VS. EDD RATIO}
\end{figure}

To determine if the nuclear properties (Section \ref{sec:modeling}) of
the VLBA detected AGN also depends on Eddington ratio, we plotted the
1.4 GHz Compactness (Equation \ref{eqn:compactness}) and compact
spectral index $\alpha_{\rm compact}$ of their nuclear radio emission
(Table \ref{Extended Table 5}) as a function of $r_{\rm Edd}$ -- shown
in Figure \ref{VS. EDD RATIO}.  We find that the 1.4 GHz compactness
of nuclear radio sources decreases as the Eddington ratio increases
(Figure \ref{VS. EDD RATIO}; {\it left}), with compact regions
responsible for $\sim80\%-100\%$ of the total 1.4 GHz nuclear emission
in the three low Eddington ratio RQQs detected in our VLBA
observations, but only $\sim30\% - 60\%$ of the emission from the
nuclear radio sources detected in the high Eddington ratio RQQs.  We
also find differences between the 1.4 - 4.8 GHz spectral index
$\alpha_{\rm compact}$ of compact regions within the nucleus of low
and high Eddingtion RQQs. Low Eddington ratio RQQs all have flat
spectra, as does the second component observed from high Eddington
ratio RQQ PG 0050+124 (PG 0050+124-C2; Fig.\ \ref{multi_comp}).
However, the first component observed from PG 0050+124 (PG
0050+124-C1; Fig.\  \ref{multi_comp}) and the other high Eddington
ratio RQQ detection both have extremely steep radio spectra.  The
observed difference in 1.4 GHz compactness and radio spectral index of
the nuclear emission of low and high Eddington RQQs suggests a
difference in physical origin. 

Using the measured compactness and spectral indices of the nuclear
radio emission, we can determine if it originates in the innermost
accretion disk, outflow, or a mixture of both.  As shown in the right
panel of Fig.\ \ref{VS. EDD RATIO}, the observed compact nature and
flat spectrum ($\alpha \sim 0$) of the nuclear emission from the
detected low Eddington ratio RQQs -- PG 2304+042, PG 1149-110, and PG
0052+251 -- suggests this emission primarily originates from the
innermost regions of the accretion disk.  This is further supported by
the relatively small offsets between the radio emission in these
sources and the optical center of the galaxy.

The diffuse 1.4 GHz emission (low compactness) from the second
component in high Eddington RQQ PG 0050+124 (PG 0050+124-C2;
Fig.\ \ref{multi_comp}) is characteristic of an outflow, but its
compact (4.8 GHz) core has a flatter radio spectrum than expected for
such emission.  This suggests that C2 may be the location of the SMBH,
with the flat-spectrum emission from the disk corona embedded within
the more diffuse, steep-spectrum, radio emission generated by the
outflow powered by this AGN (e.g., \citealt{an10, pushkarev12}).  An
alternate possibility is that C2 is a flat spectrum component of this
larger scale outflow (e.g., \citealt{hovatta14}).  Future VLBI
observations phase centered on a more precise optical position (e.g.,
from the {\it GAIA} Data Release 3; \citealt{gdr3}) could distinguish
between these possibilities, though the relatively large ($\sim60~{\rm
  mas}$) offset from the current position favors the latter.

Finally, the diffuse nature (low 1.4 GHz compactness) and steep radio
spectra ($\alpha \lesssim -1$) of the nuclear emission of high
Eddington ratio RQQ PG 1612+261 and the first component in high
Eddington RQQ PG 0050+124 (PG 0050+124-C1; Fig.\ \ref{multi_comp})
suggesting they are both produced by outflows.   

\section{Summary and Conclusions}
\label{sec:conclusions}
In this paper, we present the results of 1.4 and 4.8 GHz VLBA
observations of eight PG RQQs spanning a wide range of Eddington
ratios ($r_{\rm Edd} \sim 0.03 - 3$; Table \ref{Extended Table 1}).
Our analysis indicates that the radio properties of the observed RQQs
strongly depend on their Eddington ratio: as $r_{\rm Edd}$ increases,
a smaller fraction of its total radio emission is generated in its
nucleus, and its radio nuclear emission becomes increasingly diffuse
and has a steeper radio spectrum. Furthermore, these differences are
indicative of changes in the physical origin of the radio emission
from these RQQs.  At low Eddington ratios, the innermost accretion
disk (i.e., the disk corona and/or jet base) is primarily responsible
for the total radio emission. As the Eddington ratio increases, the
size and relative contribution of outflows increases until they
dominate the emission at the highest Eddington ratios.  Such a
dependence is not observed from radio-loud quasars, but similar
results are obtained in studies of other manifestations of radio-quiet
AGN, e.g. Narrow-Line Seyfert 1s (e.g., \citealt{doi15, yao21}).
Therefore, our results likely hold for the broader population of
radio-quiet AGN.  

Furthermore, a similar behavior is observed from stellar-mass black
hole binaries (BHBs), whose radio properties depend strongly on the
``spectral state" (defined using a combination of Eddington ratio and
X-ray spectrum) of these systems (see recent review by
\citealt{gallo10}).  The radio emission of BHBs in the ``low/hard"
spectral state, systems with an Eddington ratio $\lesssim0.03$
(e.g. \citealt{dunn10}), is typically dominated by an extremely
compact, flat spectrum radio source -- similar to what we observe for
PG 2304+042 (which has a comparable Eddington ratio).  As the
Eddington ratio of a BHB increases, the contribution of the flat
spectrum, compact core decreases, and their radio emission is
increasingly dominated by optically thin synchrotron radiation
generated by an outflow or weak jet (e.g., \citealt{gallo10}).  This
is consistent with the increasing size, and steepening spectrum, of
the nuclear radio emission with Eddington ratio we observe from the
RQQs in our sample.  Lastly, the transition of BHBs into the
``high/soft" spectral state, when they have the highest accretion
rates, is often accompanied by the ejection of radio-emitting
optically thin plasmons from the accretion disk -- similar to the
steep spectrum components observed in the nuclear regions of PG
1612+261 and PG 0050+124, the two high Eddington ratio RQQs in our
sample detected by the VLBA.  This correspondence between the nuclear
radio emission of RQQs and stellar-mass black hole X-ray binaries
across a wide range of Eddington ratios (and spectral states) is
strong evidence for the universality of accretion onto black holes. 

\begin{acknowledgments}
We would like to thank the anonymous referee for useful comments.
E.B. acknowledges support by a Center of Excellence of the Israel
Science Foundation (grant no. 2752/19).  Basic research at NYU Abu
Dhabi is supported by the Executive Affairs Authority of Abu Dhabi.
AA acknowledges support by the Kawader Research Assistantship Program.
JDG acknowledges support by NYUAD Research Grant AD022.  IZ
acknowledges support by NYUAD Research Grant AD013.  The National
Radio Astronomy Observatory is a facility of the National Science
Foundation operated under cooperative agreement by Associated
Universities, Inc.  AIPS is produced and maintained by the National
Radio Astronomy Observatory, a facility of the National Science
Foundation operated under cooperative agreement by Associated
Universities, Inc.  This work made use of the Swinburne University of
Technology software correlator \citep{deller11}, developed as part of
the Australian Major National Research Facilities Programme and
operated under licence.  This research has made use of the NASA/IPAC
Extragalactic Database (NED), which is operated by the Jet Propulsion
Laboratory, California Institute of Technology, under contract with
the National Aeronautics and Space Administration.
\end{acknowledgments}

\bibliography{sn-bibliography}
\bibliographystyle{aasjournal}
\end{document}